\documentclass[twoside,twocolumn,aps,prc,superscriptaddress,nofootinbib,showpacs,floatfix]{revtex4-1}
\usepackage{amsmath,amssymb,graphicx}
\usepackage{braket,bm}
\usepackage{morefloats}
\usepackage[english]{babel}

\newcommand{\dslash}[1]{#1\!\!\!/}
\newcommand*{\NPole}{{\text{NP}}}
\newcommand*{\Pole}{{\text{P}}}
\newcommand*{\IC}{{\text{int}}}
\newcommand*{\tJ}{\tilde{J}}
\newcommand*{\tF}{\tilde{F}}
\newcommand*{\tT}{\tilde{T}}
\newcommand*{\tkappa}{\tilde{\kappa}}
\newcommand*{\lon}{{\textsc{l}}}
\newcommand*{\tra}{{\textsc{t}}}

\newcommand*{\KR}{{\textsc{kr}}}
\newcommand*{\gpi}{g_{\textit{NN}\pi}}

\newcommand*{\TO}{{\text{TO}}}
\newcommand*{\bv}[1]{\textbf{#1}}

\allowdisplaybreaks

\begin{document}

\title{Pion photoproduction in a dynamical coupled-channels model}

\author{F.~Huang}
 \email{huang@physast.uga.edu}
 \affiliation{Department of Physics and Astronomy, The University of Georgia, Athens, GA 30602, USA}

\author{M.~D\"oring}
 \affiliation{Helmholtz-Institut f\"ur Strahlen- und Kernphysik (Theorie) and  Bethe Center for Theoretical Physics, Universit\"at Bonn, Nu{\ss}allee 14-16, 53115 Bonn, Germany}

\author{H.~Haberzettl}
 \email{helmut.haberzettl@gwu.edu}
 \affiliation{Institute for Nuclear Studies and Department of Physics, The George Washington University, Washington, DC 20052, USA}

\author{J.~Haidenbauer}
 \affiliation{\mbox{Institut f{\"u}r Kernphysik and J\"ulich Center for Hadron Physics, Forschungszentrum J{\"u}lich, 52425 J{\"u}lich, Germany}}
 \affiliation{Institute for Advanced Simulations, Forschungszentrum J\"ulich, 52425 J\"ulich, Germany}

\author{C.~Hanhart}
 \affiliation{\mbox{Institut f{\"u}r Kernphysik and J\"ulich Center for Hadron Physics, Forschungszentrum J{\"u}lich, 52425 J{\"u}lich, Germany}}
 \affiliation{Institute for Advanced Simulations, Forschungszentrum J\"ulich, 52425 J\"ulich, Germany}

\author{S.~Krewald}
 \affiliation{\mbox{Institut f{\"u}r Kernphysik and J\"ulich Center for Hadron Physics, Forschungszentrum J{\"u}lich, 52425 J{\"u}lich, Germany}}
 \affiliation{Institute for Advanced Simulations, Forschungszentrum J\"ulich, 52425 J\"ulich, Germany}

\author{U.-G.~Mei\ss ner}
 \affiliation{Helmholtz-Institut f\"ur Strahlen- und Kernphysik (Theorie) and Bethe Center for Theoretical Physics, Universit\"at Bonn, Nu{\ss}allee 14-16, 53115 Bonn, Germany} \affiliation{\mbox{Institut f{\"u}r Kernphysik and J\"ulich Center for Hadron Physics, Forschungszentrum J{\"u}lich, 52425 J{\"u}lich, Germany}}

\author{K.~Nakayama}
 \email{nakayama@uga.edu}
 \affiliation{Department of Physics and Astronomy, The University of Georgia, Athens, GA 30602, USA}
 \affiliation{\mbox{Institut  f{\"u}r Kernphysik and J\"ulich Center for Hadron Physics, Forschungszentrum J{\"u}lich, 52425 J{\"u}lich, Germany}}

\date{\today}

\begin{abstract}
The charged and neutral pion photoproduction reactions are investigated in a
dynamical coupled-channels approach based on the formulation of Haberzettl,
Huang, and Nakayama [Phys.\ Rev.\ C \textbf{83}, 065502 (2011)]. The hadronic
final-state interaction is provided by the J\"ulich $\pi N$ model, which
includes the channels $\pi N$ and $\eta N$ comprising stable hadrons as well as
the effective $\pi\pi N$ channels $\pi\Delta$, $\sigma N$, and $\rho N$. This
hadronic model has been quite successful in describing $\pi N \to \pi N$
scattering for center-of-mass energies up to $1.9$ GeV. By construction, the
full pion photoproduction current satisfies the generalized Ward-Takahashi
identity and thus is gauge invariant as a matter of course. The calculated
differential cross sections and photon spin asymmetries up to 1.65 GeV
center-of-mass energy for the reactions $\gamma p\to \pi^+n$, $\gamma p\to
\pi^0p$, $\gamma n\to \pi^-p$ and $\gamma n\to \pi^0n$ are in good agreement
with the experimental data.
\end{abstract}


\maketitle

\section{Introduction}  \label{sec:introduction}

Presently, there is intense experimental effort to study the production and the
decay of baryon resonances  from threshold to invariant collision energies of
about $\sqrt{s}=2.8$ GeV, to obtain information about the non-perturbative sector of Quantum Chromodynamics, as discussed, e.g., in the recent review \cite{Klempt:2009pi}. To deduce resonance
parameters from the experimental data, one has traditionally relied on partial-wave analysis. Most of our knowledge on baryon resonance masses is due to the classic partial-wave analyses of pion-nucleon scattering by Cutkosky \cite{Cutkosky:1979fy,Cutkosky80}, H\"ohler \cite{Hoehler83,Hohler93}, and Arndt \cite{Arndt95,Arndt02,Arndt06}, and their collaborators. In the energy range under consideration, however, reaction channels other
than pion-nucleon open up, which leads to ambiguities in the analyses. Moreover,
the number of partial waves required scales with the energy. This situation
calls for other theoretical approaches. There are a few physical principles
that an analysis should respect, such as unitarity and analyticity of the
$S$-matrix, and, as a matter of course, gauge invariance if photoproduction is
considered. The various theoretical approaches to coupled-channels
problems can be grouped into three broad classes, (i) unitarized chiral
perturbation theory, (ii) dynamical coupled channel approaches, and
(iii) $K$-matrix approaches. All classes guarantee the two-body unitarity of
the $S$-matrix  by deriving the $T$-matrix from a Bethe-Salpeter or
Lippmann-Schwinger equation, formally $ T = V + V G_0 T$, where $V$ denotes the
scattering kernel defined by a set of Born diagrams, while $G_0$ stands for the
intermediate two-particle propagator. The three classes differ by the
choice of the scattering kernel $V$ and the propagator $G_0$.

(i) In photoproduction reactions, the threshold region is understood in terms
of chiral perturbation theory (ChPT) \cite{Bernard:1991rt,Bernard92,Bernard:1994gm} (for a recent comprehensive review see \cite{Bernard:2007zu}). Unitarization of the interaction allows
one to extend the applicability into the resonance region. Such {\it
chiral unitary approaches} respect chiral symmetry, a fundamental property of
the strong interaction. Moreover, the scattering kernel $V$ is obtained within
a systematic counting scheme which limits the number of admissible diagrams
according to the order of the chiral expansion. A scheme to gauge-invariantly
couple the photon to the unitarized amplitude was developed in
Ref.~\cite{Borasoy:2005zg} and applied to kaon and eta photoproduction
\cite{Borasoy:2007ku,Ruic:2011wf}. See also Ref.~\cite{DN10} for a
gauge-invariant unitary framework in the context of the recently
discovered structure in $\eta$ photoproduction on the neutron. For earlier works on photoprocesses in the chiral unitary framework, see, e.g., Refs.~\cite{Kaiser:1996js,Nacher:1999ni,Marco:1999df}.

Chiral unitary approaches usually concentrate on $S$-waves although some
authors consider higher-order terms in the scattering kernel, thus allowing
them to study the $S$ and $P$ partial waves within this approach
\cite{MO00,Meissner00}. A unitary coupled-channels model was developed in Ref.~\cite{Gasparyan10} where the partial-wave amplitudes for the $\gamma N$ and $\pi N$ states are obtained by analytic extrapolations of the subthreshold reaction amplitudes computed in ChPT. Chiral unitary approaches respect the analyticity of the $T$-matrix by keeping both the real and the imaginary parts
of the two-particle propagator $G_0$. An interesting physical consequence of analyticity is the possibility to generate bound states or resonances by meson-baryon dynamics alone. The $N^*(1535)$, $\Delta^*(1700)$, and other resonances have been claimed to be dynamically generated \cite{KSW95,KL04,SOV05,Doring:2007rz,Jido:2007sm,Bruns:2010sv}. While chiral unitary approaches are certainly the most elegant ones theoretically, their actual applications to the analysis of data and the study of resonances has been limited to low partial waves and an energy range well below
2 GeV.

(ii) Dynamical coupled-channels approaches employ effective meson-baryon
Lagrangians to define the scattering kernel $V$ of the Bethe-Salpeter equation,
giving up the  chiral counting scheme for $V$. Analyticity of the $T$-matrix is
guaranteed by solving the Lippmann-Schwinger equation employing the full
two-particle propagator. There is no restriction on the partial waves, which is
essential for analyzing observables at higher energies. The scattering kernel
$V$ includes the exchange of mesons in the $t$-channel and the exchange of
baryons in the $u$-channel and therefore generates correlations between different partial waves. As such approaches are a Lagrangian based, SU(3) symmetry allows also to correlate different reaction channels. Those correlations introduce an energy dependence to the non-resonant background
which is not due to the presence of $s$-channel resonances. In principle, the method can generate baryon resonances dynamically, a feature shared with the chiral unitary approach, but in actual calculations, in most cases resonant $s$-channel driving terms are required for quantitative rendering of the data because the partial-wave correlations put tight constraints on the free parameters of the model that inhibits to some degree the possibility of generating resonances dynamically. Still, it should be stressed that due to the non-resonant background, only a small number of $s$-channel resonances is required which should prevent the approach from claiming spurious resonances in the data \cite{Ceci:2011ae}. As dynamical coupled-channels approaches respect analyticity, the analytic continuation of the amplitudes is possible and provides the poles and the residues of the $S$-matrix on the various Riemann sheets. As field-theoretical quantities, the poles of the $S$-matrix do not suffer from ambiguities, as for
example Breit-Wigner fits do, and thus provide a more model-insensitive way of characterizing the baryon resonances. At this point, we would like to stress the importance of a dynamical treatment of three-body cuts, as they appear, e.g., in intermediate $\pi\pi N$ states. In Ref.~\cite{Ceci:2011ae} it has been shown that the resulting branch points in the complex plane are important for the extraction of the baryon spectrum; in models that do not contain these points, they may be simulated by poles, leading to potentially erroneous results.

An actively developed dynamical coupled-channels approach is the J\"ulich model
\cite{Schutz:1994ue,Schutz98,Krehl00,Gasparyan03,RDSHM10} that includes the
$\pi N$, $\eta N$, $\sigma N$, $\rho N$, and $\pi\Delta$ hadronic channels.
This approach provides the hadronic part of the interaction in the present
study, as specified below. Quite recently, the J\"ulich model has incorporated
the $K\Lambda$ and $K\Sigma$ channels \cite{RDSHM10}, using SU(3) symmetry, and
a global fit to the corresponding experimental data is in progress. A
feature in the J\"ulich model, unique among modern dynamical coupled-channels approaches, is the treatment of the important $t$-channel exchanges with $\rho$ and $\sigma$ quantum numbers: those are largely fixed using crossing-symmetry and dispersive techniques from
$\bar NN\to\pi\pi$ data \cite{Schutz:1994ue}, thus drastically reducing the
model dependence at this point. Progress in a different direction has  been
recently achieved in Ref.~\cite{Doring:2011ip} where it has been shown for the
J\"ulich and similar models that the approach is suited to analyze upcoming
lattice data on the baryon spectrum. A scheme has been developed to address
finite-volume effects and lattice levels could be predicted.

Among the various groups developing dynamical coupled-channels approaches \cite{Surya:1995ur,Mainz} is the EBAC group \cite{Matsuyama07}. Their approach includes explicit $\pi\pi N$ contributions
in an approximate manner in an effort to satisfy some aspects of three-body
unitarity. Like most coupled-channels models, it includes the $\gamma N$ channel in the one-photon
approximation \cite{Diaz08}. It is not gauge invariant, however. The EBAC
approach has been applied extensively \cite{Diaz08,Diaz09,Kamano09} in the
analysis of pion photo- and electroproduction data.

(iii) Another class of approaches used to analyze pion- and photon-induced reactions are $K$-matrix models. Dynamical coupled-channels approaches require a price for what is delivered: one has to solve coupled integral equations. The technical effort can be reduced by approximating the two-body propagator $G_0$ in the Lippmann-Schwinger equation: omitting the real part of the two-body propagator, one reduces the integral equations to a set of algebraic equations. In such $K$-matrix approaches, unitarity is still respected due to the presence of the imaginary part of $G_0$, but analyticity is lost. As a consequence of the approximation, only on-shell intermediate
states are taken into account when solving the scattering equation,  while the
principal-value (dispersive) parts of the scattering equation are neglected,
which  suppresses the full contributions of the virtual two-body intermediate
states and  in general results in a reduction of the strengths of multiple-scattering contributions. The resonance parameters may compensate for this approximation. Due to its technical simplicity and flexibility, the $K$-matrix approach has made possible the quantitative reproduction of a large body of experimental data. Recent applications of the method can be found in Refs.~\cite{Arndt02,Arndt06,Feuster99,Penner02,KVI,BonnGatchina,Anisovich:2010an}.
A variation of the standard $K$-matrix approach (unitary isobar model) was
developed by the Mainz group \cite{Drechsel:2007if,Tiator:2011pw}.

In the present work, we present a gauge-invariant treatment of pion
photoproduction. A feasibility study of this reaction employing the J\"ulich
$\pi N$ model was presented in Ref.~\cite{Haberzettl06}. The corresponding differential
cross sections for neutral and charged pion photoproduction reactions were
found to be in reasonable agreement with the data up to the total
center-of-mass (c.m.) energy of $W=1.25$ GeV. In this study, we will extend and
refine the calculations of Ref.~\cite{Haberzettl06} employing a more recent
version of the J\"ulich model \cite{Gasparyan03} for the hadronic part of the
amplitude, and combine it with a novel way of solving for the photoproduction
current in a gauge-invariant manner \cite{HHN2011}.

The J\"ulich $\pi N$  model \cite{Schutz98,Krehl00,Gasparyan03} is based on
time-ordered perturbation theory (TOPT) \cite{Schweber}. It is a
coupled-channels meson-exchange model including the $\pi N$ and $\eta N$ channels as well as the $\pi\Delta$, $\sigma N$, and $\rho N$ effective channels which implicitly account for the resonant part of the $\pi\pi N$ channel. The interaction kernel corresponding to the $t$- and $u$-channel
diagrams is constructed based on the (effective) chiral Lagrangians of Wess and
Zumino \cite{Wess67,Meissner:1987ge}, taking into account the corresponding
lowest-order non-vanishing terms. This Lagrangian is supplemented by additional
terms for the coupling of $\Delta$, $\omega$, $\eta$, $a_0$, and $\sigma$
\cite{Krehl00,Gasparyan03}. For the $s$-channel diagrams, apart from the bare
nucleon pole contribution which is renormalized by the coupling to the $\pi N$
continuum state to reproduce the physical nucleon, the interaction kernel
includes eight genuine resonances, namely $S_{11}(1535)$, $S_{11}(1650)$,
$S_{31}(1620)$, $P_{31}(1910)$, $P_{13}(1720)$, $D_{13}(1520)$, $P_{33}(1232)$,
and $D_{33}(1700)$. The bare genuine resonances get their dressed masses and
widths from the re-scattering of baryon-meson continuum states; the
$P_{11}(1440)$ (Roper) resonance appears as a dynamically generated resonance
due to the strong interaction within and between the $\pi N$ and $\sigma N$
channels. This hadronic model has been quite successful in reproducing the $\pi
N$ partial-wave amplitudes with total angular momentum $J=1/2$ and $3/2$ up to
a c.m.\ energy of $1.9$ GeV. Recently, also the pole positions and residues of
the resonances in this model have been extracted \cite{Doring09L,Doring09}.

As alluded to above, in Ref.~\cite{Haberzettl06} a dynamical coupled-channels
model for pseudoscalar meson photoproduction based on the field-theory approach
of Haberzettl \cite{Haberzettl97} was introduced in conjunction with the
J\"ulich hadronic coupled-channels model \cite{Krehl00}. For the present
application, however, we will use the reformulation of Haberzettl, Huang, and
Nakayama \cite{HHN2011}, which differs from the original approach of
Haberzettl, Nakayama, and Krewald \cite{Haberzettl06} in some essential aspects
that provide several practical advantages, as we shall explain in more detail
in Sec.~\ref{sec:formalism}.

In general, the present photoproduction approach is distinguished from most
existing dynamical models by the fact that it satisfies the generalized
Ward-Takahashi identity for the production current \cite{Kazes59,Haberzettl97}
that ensures its full gauge invariance. By contrast, the vast majority of
existing dynamical models at best provide only a conserved current but are not
truly gauge invariant in their internal dynamics (see subsequent paragraph). An
exception to this is the model of Refs.~\cite{Pascalutsa04,Caia04,Caia05},
where gauge invariance is implemented following the prescription of Gross and
Riska \cite{Gross}. See also Ref.~\cite{AA95}, where the issue of gauge
invariance in pion photoproduction is discussed.
An alternative method for achieving gauge invariance was developed in
Ref.~\cite{Kvinikhidze09}; however, it has not been applied in practical
calculations so far.

The present approach \cite{Haberzettl97,Haberzettl06,HHN2011} is based on a
fully \textit{microscopic} (i.e., local) implementation of gauge invariance in
which each electromagnetic current contribution associated with an internal
subprocess of the reaction satisfies its own individual \textit{off-shell}
Ward-Takahashi identity (WTI) --- either as an ordinary WTI for a
single-particle current \cite{WTI} or as a generalized WTI for an interaction
current \cite{Kazes59,Haberzettl97} --- thus ensuring the overall gauge
invariance of the physical current matrix elements. It was emphasized in
Ref.~\cite{Haberzettl97} that this feature is essential for making each
individual contribution a consistent building block for the correct description
of the reaction dynamics of the entire process. Most importantly, these
$n$-point-current building blocks (and their associated WTIs) remain the same
whatever the dynamical context in which they appear. This is not simply a
purely theoretical issue concerning the aesthetics of the formulation, but this
consistency requirement across all possible reactions has immediate practical
consequences. This was most clearly demonstrated in the recent study of the
$NN$ bremsstrahlung reaction \cite{HN10,NH09} whose fully gauge-invariant
reaction amplitude was obtained within the same field-theory framework
\cite{Haberzettl97} that provides the basis for the present formulation of pion
photoproduction \cite{HHN2011}. Specifically, it was shown in
Refs.~\cite{HN10,NH09} that essential aspects of the bremsstrahlung process can
be understood as time-reversed meson photoproduction processes $NM\to N\gamma$
(where $M$ stands for a meson) whose dynamical details, therefore, can be
described by the current building blocks from the corresponding
meson-production formulation. The most important building block in this context
turned out to be the four-point interaction current for the subprocess $NM\to
N\gamma$ whose \textit{gauge-invariant} construction in terms of its
corresponding interaction-current WTI provided the contribution necessary to
resolve the longstanding discrepancy of nearly a decade between the
high-precision KVI data \cite{KVI02} and the then existing models of the $NN$
bremsstrahlung reaction.

In the present work, going beyond the feasibility study given in
Ref.~\cite{Haberzettl06}, we carry out an extended and more quantitative
calculation of the $\gamma p\to \pi^+n$, $\gamma p\to \pi^0p$, $\gamma n\to
\pi^-p$ and $\gamma n\to \pi^0n$ reactions using the input of the J\"ulich $\pi
N$ model of Ref.~\cite{Gasparyan03} to describe the hadronic final-state
interactions (FSI) of the reactions. The pion photoproduction current itself is
calculated using the novel formulation of Ref.~\cite{HHN2011} whose details are
given in the subsequent section. To keep the equations manageable numerically,
as a first step toward a more complete calculation, we explicitly include only
the $\eta N$ and $\pi\Delta$ channels and the basic $\pi N$ channel in the FSI
loop integration. Of course, the corresponding meson-baryon to meson-baryon
$T$-matrices contain the information of all hadronic channels that comprise the
J\"ulich model. We calculate cross sections as well as beam asymmetries up to
the total c.m.\ energy of $W=1.65$ GeV.

The present paper is organized as follows. In Sec.~\ref{sec:formalism}, we
briefly present our covariant formalism for pion photoproduction. In
Sec.~\ref{sec:Juelich}, we explain how the J\"ulich hadronic model --- which is
based on TOPT --- has been reformulated and matched to the present covariant
formalism. In Sec.~\ref{sec:furtherdetail}, details of additional
approximations made in our photoproduction model are presented. Our results for
both the neutral and charged pion photoproduction reactions are presented in Sec.~\ref{sec:results}, where some discussion is presented as well. Section~\ref{sec:uncertainties} contributes to the discussion about the theoretical uncertainties of our results. Finally, the summary and discussion of future developments are given in Sec.~\ref{sec:summary}. Some details of the present model (interaction Lagrangians, form factors and the corresponding parameter values) are given in the Appendix.

\section{Formalism} \label{sec:formalism}

\begin{figure}[t!]
  \includegraphics[width=\columnwidth,clip=]{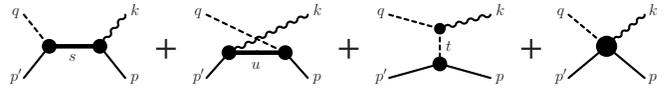}
  \caption{\label{fig:gammaN}
  Generic structure of the pion photoproduction current $M^\mu$ for $\gamma N
  \to \pi N$ according to Eq.~(\ref{eq:MmuGeneric}). Time proceeds from right
  to left. Nucleons and pions are depicted by solid and dashed lines,
  respectively, and the photon is shown as a wavy line. The four-momenta shown
  at the external lines are those of the respective particles used in this
  work. The first three diagrams comprising class-A diagrams \cite{GellMann54}
  depict the $s$-, $u$-, and $t$-channel pole diagrams $M^\mu_s$, $M^\mu_u$,
  and $M^\mu_t$, respectively, where $s$, $u$, and $t$ are the respective
  Mandelstam variables of the internally exchanged particles, as indicated. The
  last diagram (called class B \cite{GellMann54}) shows the contact-type
  interaction current $M^\mu_\IC$.}
\end{figure}

The basic structure of pion photoproduction current $M^\mu$ seems fairly simple
topologically. As shown in Fig.~\ref{fig:gammaN}, it is comprised of the three
$s$-, $u$-, and $t$-channel contributions  $M^\mu_s$, $M^\mu_u$, and $M^\mu_t$,
respectively, where the photon is attached to the external legs of the basic
$\pi NN$ vertex and one diagram where the photon interacts with the interior of
the vertex (correspondingly called the interaction current $M^\mu_\IC$), i.e.,
\begin{equation}
  M^\mu= M^\mu_s +M^\mu_u+M^\mu_t+M^\mu_\IC~.
  \label{eq:MmuGeneric}
\end{equation}
In general, all four current contributions here are fully dressed.  Following
Ref.~\cite{GellMann54}, the first three (pole-type) diagrams are usually called
class-A diagrams and the last (non-pole) diagram is of class B. This simple
topological structure is also reflected by the corresponding basic tree-level
Feynman diagrams for the process when reducing the full complexity of the
current to using bare propagators and vertices only. In full detail, however,
the microscopic dynamics of the production process is much, much more complex
because of dressing effects of the hadron propagators and vertices. Moreover,
since particle number is not conserved (e.g., internally any number of mesons
can partake in the process), the full process is highly non-linear as a matter
of course.

Comprehensive theoretical formulations of photoproduction processes must be
able to incorporate these dynamical complications at least in principle. In
addition, the corresponding production current $M^\mu$ must obey gauge
invariance as the manifestation of U(1) symmetry which is of fundamental
importance for any photoprocess because it provides a conserved current and
thus implies charge conservation. For the pion production current $M^\mu$, in
particular, gauge invariance is formulated in terms of the generalized
Ward-Takahashi identity \cite{Kazes59,Haberzettl97},
\begin{align}
  k_\mu M^\mu &= -F_s S(p+k) Q_i S^{-1}(p)
  \nonumber\\
  &\qquad\mbox{}
  + S^{-1}(p')Q_f S(p'-k) F_u
  \nonumber\\
  &\qquad\quad \mbox{}
  + \Delta^{-1}_\pi(q) Q_\pi \Delta_\pi(q-k) F_t~,
  \label{eq:gWTI}
\end{align}
where the four-momenta are those shown in Fig.~\ref{fig:gammaN}. The vertices
$F_x$ here correspond to the (fully dressed) $\pi NN$ vertices $F$ in the
respective kinematic situations corresponding to the Mandelstam variables
$x=s,u,t$, as shown in Fig.~\ref{fig:gammaN}. The propagators for the nucleon
and pion are denoted by $S$ and $\Delta_\pi$, respectively, and the charge
operators for the initial and final nucleon and for the outgoing pion are
$Q_i$, $Q_f$, and $Q_\pi$, respectively. Obviously, this expression vanishes
for on-shell hadrons and thus provides a conserved current. This
\textit{off-shell} formulation of gauge invariance, however, goes beyond that
by providing a \textit{local} constraint on the gauge invariance of the
photoproduction current that is similar to requiring the usual Ward-Takahashi
identity for the single-particle currents \cite{WTI}, which for the nucleon
reads
\begin{equation}
  k_\mu J^\mu = S^{-1}(p+k)Q_N -Q_N S^{-1}(p)~,
  \label{eq:NWTI}
\end{equation}
where $Q_N$ is the nucleon's generic charge operator. Both requirements
(\ref{eq:gWTI}) and (\ref{eq:NWTI}) (and its analog for the pion) are essential
for the internal consistency of microscopic formulations of
photoprocesses.\footnote{As mentioned in the Introduction, a particularly
striking example of this, with immediate practical consequences for the
description of experimental data, was recently discussed for $NN$
bremsstrahlung \cite{NH09,HN10}.}

%=======================================================
%=======================================================

One can easily show that assuming the validity of the usual WTI (\ref{eq:NWTI})
for the electromagnetic nucleon current $J^\mu$ and its analog for the pion
current, the generalized WTI (\ref{eq:gWTI}) implies
\begin{equation}
  k_\mu\,M^\mu_\IC = -\tF_s e_i + \tF_u e_f + \tF_t e_\pi~,
  \label{eq:gWTIint}
\end{equation}
where the $\tF_x$ are the vertices $F_x$ of (\ref{eq:gWTI}) stripped of their
isospin operators $\tau$ that now appear in
\begin{equation}
  e_i = \tau Q_i~,\qquad
  e_f = Q_f \tau~,\quad\text{and}\quad
  e_\pi = Q_\pi \tau
\end{equation}
which are the charges for all external hadron legs in an appropriate isospin
basis (with all corresponding indices and summations suppressed). In other
words, the relation
\begin{equation}
  e_i =e_f+e_\pi
\end{equation}
describes charge conservation for the pion photoproduction process. With the
single-particle WTIs for nucleons and pions given, Eq.~(\ref{eq:gWTIint}) is
completely equivalent to the generalized WTI (\ref{eq:gWTI}). It is this
 Eq.~(\ref{eq:gWTIint}) for the contact-type interaction current $M^\mu_\IC$,
 in particular, that is being exploited here for the purposes of preserving the
 overall gauge invariance of the production current.

\subsection{Dyson-Schwinger framework}

The complete (i.e., fully dressed, non-linear, and gauge-invariant) structure
of pion photoproduction was described by Haberzettl \cite{Haberzettl97} within
a covariant field-theoretical Dyson-Schwinger framework. In practice, however,
the complexity of the full formalism needs to be truncated at some level to
make it numerically tractable. Since doing so invariably leads to a violation
of gauge invariance, one must find prescriptions to restore this fundamental
symmetry. And one should do so in a manner that incorporates as many of the
original reaction mechanisms as possible. Particularly relevant in this context
are the hadronic final-state interactions of the outgoing meson-nucleon states
since these form essential parts of the dynamical content of the interaction
current $M^\mu_\IC$ \cite{Haberzettl97}.

In Ref.~\cite{Haberzettl06}, it was shown how to approximate the full formalism
in a manner that includes the full hadronic final-state interactions while at
the same time preserving the gauge invariance by reproducing the generalized
Ward-Takahashi identity (\ref{eq:gWTI}) for the production current. In the
present paper, we will follow the variant of the procedure of
Ref.~\cite{Haberzettl06} put forward recently in Ref.~\cite{HHN2011}. We
emphasize that without any approximation, i.e., as full Dyson-Schwinger
formulations,  the respective results of
Refs.~\cite{Haberzettl06,Haberzettl97,HHN2011} are completely equivalent.
However, the features of Ref.~\cite{HHN2011} turn out to be particularly well
suited for approximating the dressing effects of the electromagnetic nucleon
current $J^\mu$ in a way that is reciprocally consistent with the production
current $M^\mu$ itself while at the same time preserving local gauge
invariance. Here, we will recapitulate the details of the approach given in
Ref.~\cite{HHN2011} only insofar as it is necessary for the present
application. To this end, we make extensive use of diagrammatic
representations. Full details and derivations can be found in
Ref.~\cite{HHN2011}, and in Refs.~\cite{Haberzettl97,Haberzettl06}.

%=======================================================
\begin{figure}[t!]
 \includegraphics[width=\columnwidth,clip=]{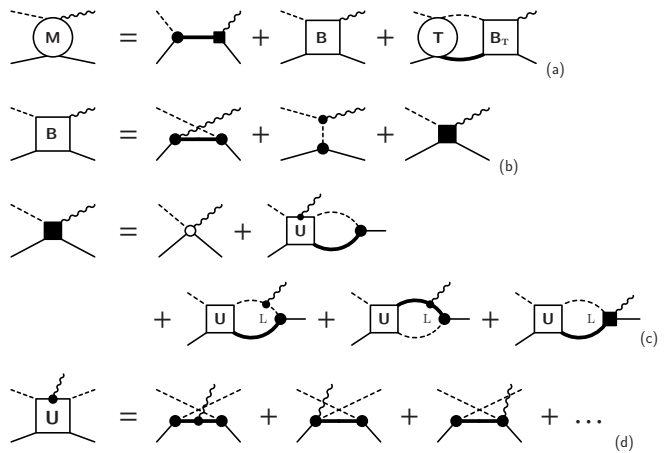}
 \caption{\label{fig:MwT}
  Microscopic structure of the photoproduction process according to
  Ref.~\cite{HHN2011}. The line styles are the same as in
  Fig.~\ref{fig:gammaN}, with thick solid nucleon lines depicting the
  propagation of dressed intermediate nucleons. Solid and open circles indicate
  dressed and bare vertices, respectively. (a) The full current $M^\mu$ with
  its FSI contribution written as a loop integration over the full $\pi N$
  $T$-matrix. (b) Details of the Born-type current $B^\mu$ of
  Eq.~(\ref{eq:BornCurrent}) that comprises the $u$ and $t$-channel
  contributions $M^\mu_u$ and $M^\mu_t$, and the contact-type four-point
  current $M^\mu_c$ depicted as a solid square whose details in turn are shown
  in part (c). The first term here is the Kroll-Ruderman contact term
  $m^\mu_\KR$. The various hadronic boxes labeled $U$ depict the non-polar $\pi
  N$ irreducible driving term of the $\pi N$ $T$-matrix, as shown in
  Fig.~\ref{fig:TXU}(c). The particular five-point current $U^\mu$ appearing in
  the first loop arises from attaching a photon to $U$; its lowest order is
  shown in part (d). The four-point currents appearing here correspond to the
  \textit{full} interaction current $M^\mu_\IC$ of Fig.~\ref{fig:gammaN}. The
  box labeled $B_\tra$ in (a) and the loop integrations labeled L in (c)
  indicate restrictions of the corresponding $B^\mu$ contributions to
  transverse and longitudinal pieces, respectively (see text). The details of
  the full nucleon current $J^\mu$ (solid circle) and its auxiliary $s$-channel
  restriction $\tJ^\mu_s$ (solid square) are depicted in
  Fig.~\ref{fig:NcurrentC}.}
\end{figure}
%=======================================================

Note that on the hadronic side, for simplicity, we will speak here most of the
time explicitly only of pions and nucleons. However, these particles are to be
taken as representatives for all mesons and baryons, respectively, that take
part in a particular application. Correspondingly, any explicit equation
appearing here for the $\pi N$ channel is to be read as a matrix equation that
couples all hadronic meson-baryon channels once the full complexity of the
reaction dynamics is turned on.

%=======================================================
\begin{figure*}[t]
  \includegraphics[width=.75\textwidth,clip=]{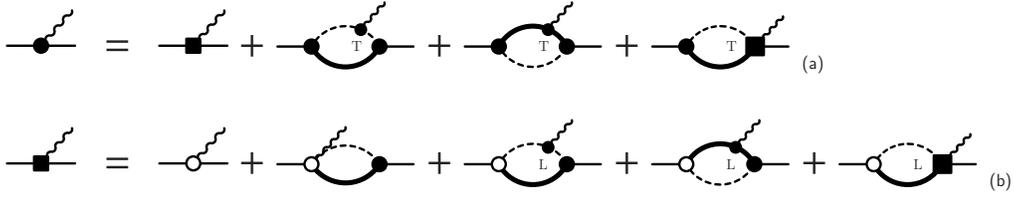}
  \caption{\label{fig:NcurrentC}
  (a) The dressed electromagnetic nucleon current $J^\mu$, with the first term
  on the right-hand side (depicted as a solid-square three-point current)
  corresponding to $\tJ^\mu_s$ appearing in Eq.~\ref{eq:MmuWithT}. (b) The
  details of $\tJ^\mu_s$. The first term (with open circle) is the bare current
  and the second term contains the Kroll-Ruderman term (open-circle four-point
  current) integrated over by the dressed $\pi NN$ vertex (solid circle; cf.\
  Fig.~\ref{fig:hadrondressing}). The remaining loop contributions only
  contribute longitudinally (as indicated by the index L). By contrast, the
  loops in (a) only contribute transversally, as indicated by the letter T. The
  solid-square contact-type four-point current $M^\mu_c$ in the loops at the
  end of both lines is given in Fig.~\ref{fig:MwT}(c).}
\end{figure*}
%=======================================================

\begin{figure}[t!]
  \includegraphics[width=.75\columnwidth,clip=]{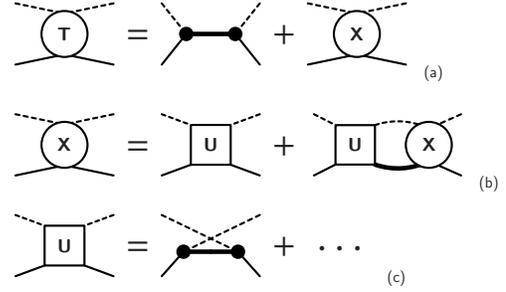}
  \caption{\label{fig:TXU}
  (a) Representing the full $T$-matrix for $\pi N$ scattering via its (fully
  dressed) $s$-channel pole part and its non-polar contribution $X$, according
  to Eq.~(\ref{eq:TXsplit}). (b) The Bethe-Salpeter scattering equation for the
  non-pole amplitude $X$, Eq.~(\ref{eq:XBSeq}). Its driving term $U$ comprises
  \textit{all} $\pi N$ irreducible mechanisms. The corresponding lowest-order
  contribution is the $u$-channel exchange shown explicitly in (c). Attaching a
  photon to this diagram provides the five-point current given in
  Fig.~\ref{fig:MwT}(d).}
\end{figure}

\begin{figure}[t!]\centering
\includegraphics[width=.7\columnwidth,clip=]{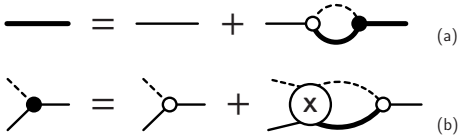}
  \caption{\label{fig:hadrondressing}
  Dressing mechanisms for (a) the nucleon propagator $S$ in terms of the bare propagator $S_0$ (thin line) and the
  self-energy loop and (b) the $\pi NN$ vertex $F$ according to
  Eq.~(\ref{eq:dressedF}). Dressed and undressed vertices are shown as solid
  and open circles, respectively.}
\end{figure}

Following Ref.~\cite{HHN2011}, the full pion photoproduction current $M^\mu$
in the one-photon approximation\footnote{This is not a serious
limitation since higher-order contributions are suppressed by $e^2\approx
1/137$. For most if not all practical purposes, therefore, the one-photon
approximation is perfectly adequate.} can be written as
\begin{equation}
M^\mu = F_s S \tilde{J}^\mu_s + B^\mu + TG_0 B^\mu_\tra~,
\label{eq:MmuWithT}
\end{equation}
with the non-pole Born-type current given by
\begin{equation}
  B^\mu =M^\mu_u +M^\mu_t +M^\mu_c~.
  \label{eq:BornCurrent}
\end{equation}
These equations are depicted diagrammatically in Fig.~\ref{fig:MwT}. The (fully
dressed) $u$- and $t$-channel contributions $M^\mu_u$ and $M^\mu_t$ of
Eq.~(\ref{eq:MmuGeneric}) appear here unchanged, but the $s$-channel current
$M^\mu_s$ and the interaction current $M^\mu_\IC$ are now spread out over the
remaining terms. The first term, $F_s S \tilde{J}^\mu_s$, contains part of the
$s$-channel pole contribution via the fully dressed nucleon propagator $S$,
however, with an electromagnetic current contribution $\tJ^\mu_s$ for the
nucleon that is only part of the full nucleon current $J^\mu$. The situation is
depicted in Fig.~\ref{fig:NcurrentC} and will be discussed in more detail
below. The contact term $M^\mu_c$ [see Fig.~\ref{fig:MwT}(c)] contains part of
the full interaction current $M^\mu_\IC$. The remaining pieces for both
$M^\mu_s$ and $M^\mu_\IC$ come from the $TG_0B^\mu_\tra$ term which describes
the $\pi N$ final-state interaction mediated by the loop integration over the
full $\pi N$ $T$-matrix, where $G_0$ is the intermediate propagator of the free
$\pi N$ pair within the loop. Note that only the \textit{transverse} parts of
the contributions of these loop integrations are to be taken into account here,
as indicated by the index T.\footnote{For definiteness, since transverse parts
are not unique,
  throughout this work here the index T on a current $j^\mu$ indicates
  $j^\mu_\tra=\varepsilon^\mu \varepsilon_\nu j^\nu/\varepsilon^2$ where
  $\varepsilon^\mu$ is the (transverse) photon polarization vector. The
  splitting into longitudinal and transverse parts satisfies
  $j^\mu=j^\mu_\lon+j^\mu_\tra$.}
 The respective missing pieces for $M^\mu_s$ and $M^\mu_\IC$
arise from splitting the full $T$ into an $s$-channel pole part and a non-pole
part according to
\begin{equation}
  T = \ket{F} S \bra{F} + X~,
  \label{eq:TXsplit}
\end{equation}
where the first term contains the $s$-channel nucleon pole via the dressed
nucleon propagator $S$ and $X$ denotes the remaining non-pole part.\footnote{We
  follow here the notation of Ref.~\cite{Haberzettl97}. In other words, for
  notational simplicity, we do not employ the often-used notation $T^\Pole$ and
  $T^\NPole$ for the pole and non-pole contributions of $T$, respectively,
  since this notation tends to make the equations difficult to parse. For
  similar reasons, to avoid the extensive use of adjoint daggers, we use the
  ket and bra notation $\ket{F}$ and $\bra{F}$ to denote (dressed) $\pi NN$
  vertices that describe $N\to \pi N$ and $\pi N\to N$, respectively. Within
  the context of Eq.~(\ref{eq:MmuWithT}), therefore, the $s$-channel vertex
  $F_s$ may also be written as $\ket{F}$. The bras and kets, however, are not
  to be misconstrued as Hilbert-space vectors.}
This splitting is depicted diagrammatically in Fig.~\ref{fig:TXU} together with
the Bethe-Salpeter equation for the non-pole amplitude,
\begin{equation}
  X=U+UG_0 X~,
  \label{eq:XBSeq}
\end{equation}
driven by the (fully dressed) non-pole $\pi N$ irreducible mechanisms subsumed
in $U$ whose lowest-order contribution is the $u$-channel exchange shown in
Fig.~\ref{fig:TXU}(c). The $F$'s here are the fully dressed $\pi NN$ vertices
related to the bare vertex $f$ by
\begin{equation}
  \ket{F} = \ket{f} + XG_0 \ket{f}~.
  \label{eq:dressedF}
\end{equation}
This dressed vertex is represented diagrammatically in
Fig.~\ref{fig:hadrondressing} together with the dressing mechanism for the
nucleon propagator
\begin{equation}
  S=S_0 + S_0 \Sigma S
  \label{eq:dressedS}
\end{equation}
given in terms of the bare propagator $S_0$  and the nucleon self-energy loop
$\Sigma$.

%=======================================================
%=======================================================

%=======================================================
%=======================================================

The entire set of equations summarized by Figs.~\ref{fig:MwT} through
\ref{fig:hadrondressing} provides a complete solution of the problem by forming
a tower of coupled non-linear Dyson-Schwinger-type equations that can only be
solved iteratively. Viewing this coupled set in its entirety for a given bare
input, the splitting (\ref{eq:TXsplit}) of $T$ into pole and non-pole
contributions is unique. However, this uniqueness generally is lost once
approximations are introduced. As a consequence, the resulting individual
pieces of a particular approximation of the splitting (\ref{eq:TXsplit}) may
exhibit undesirable numerical artifacts which the $T$ matrix itself does not
possess. (This is the case, for example, for pole and non-pole parts resulting
from the J\"ulich model we use here \cite{Doring09L}.) In order to avoid such
problems to some extent,\footnote{We mention that in the one-photon
approximation, one cannot
  formulate the photoproduction process without employing a splitting of $T$
  into pole and non-pole contributions of some sort. The two opposite extremes
  of such splittings are, at one end, the Dyson-Schwinger-type splitting
  (\ref{eq:TXsplit}) and, at the other end, the splitting where the pole part
  only contains the physical pole position itself, without any dressing
  effects, and the (constant) residue at this pole, with everything else
  appearing in the non-pole part. Going beyond the one-photon approximation,
  one may perhaps be able to avoid the aforementioned problems if one makes the
  $\gamma N$ channel an integral part of the set of coupled channels thus
  producing the photoproduction current as a solution element of $T$ when
  solving the corresponding coupled-channels matrix equation (\ref{eq:TBSeq}).
  However, we have no proof of this conjecture. Moreover, preserving gauge
  invariance in that approach is much more complicated than in the present
  one.}
one may, instead of Eq.~(\ref{eq:XBSeq}), employ the Bethe-Salpeter equation
\begin{equation}
  T=V+V G_0T
  \label{eq:TBSeq}
\end{equation}
for the entire $T$-matrix, with the driving term
\begin{equation}
  V= \ket{f} S_0 \bra{f} +U~,
\end{equation}
where $f$ and $S_0$ are the bare vertex and bare propagator of
Eqs.~(\ref{eq:dressedF}) and (\ref{eq:dressedS}), respectively.

The photoproduction current of Eq.~(\ref{eq:MmuWithT}) is formulated assuming
that the hadronic scattering information enters the problem in terms of $T$,
i.e., in terms of a matrix coupled-channels equation of the generic type given
in (\ref{eq:TBSeq}). An equivalent formulation in terms of $X$
exists \cite{Haberzettl97,HHN2011}, however, since the usual hadronic
coupled-channels approaches (like the J\"ulich model we employ here) produce
full $T$-matrices as a matter of course, it is of considerable practical
advantage to utilize this information directly for the photoproduction current.
(See also the corresponding discussion at the end of Sec.~\ref{sec:Juelich}.)

\subsection{Gauge-invariant truncation of the pion photoproduction current}

As mentioned, the preceding formulation is exact, but it is also very complex
in detail and therefore, at present at least, cannot be solved in its entirety
without making some approximations to render the problem manageable
numerically. In the particular formulation \cite{HHN2011} used here, there are
two obvious places for such approximations --- the dressed nucleon current
$J^\mu$ depicted in Fig.~\ref{fig:NcurrentC} and the contact-type five-point
current $M^\mu_c$ shown in Fig.~\ref{fig:MwT}(c).

The detailed (exact) structure of the electromagnetic nucleon $J^\mu$ is given
by \cite{HHN2011}
\begin{equation}
  J^\mu = \tJ^\mu_s + \bra{F} G_0 \left(M^\mu_u+M^\mu_t+M^\mu_c\right)_\tra~,
  \label{eq:JmuDetail}
\end{equation}
as one easily reads off Fig.~\ref{fig:NcurrentC}. The currents within the loop
structure here are given entirely by pieces occurring in the pion production
current itself. As was shown in Ref.~\cite{HHN2011}, this reciprocal
consistency between the nucleon current $J^\mu$ and the production current
$M^\mu$ was essential for deriving the particular form of $M^\mu$ given in
Eq.~(\ref{eq:MmuWithT}) because the transverse loop contributions of
Eq.~(\ref{eq:JmuDetail}), in particular, survive unchanged in the loop
integration that provides the FSI contribution in Eq.~(\ref{eq:MmuWithT}). This
form, moreover, easily permits to treat both $J^\mu$ and $M^\mu$ with a
consistent set of approximations.

\subsubsection{Approximating the nucleon current $J^\mu$}\label{sec:III_Nucleon}

 First, we note that all loop integrations in Eqs.~(\ref{eq:MmuWithT}) and
(\ref{eq:JmuDetail}) contain the full nucleon current within the $u$-channel
loops. The nucleon current, thus, couples back into itself and these
contributions, therefore, are part of the non-linear hierarchy of
Dyson-Schwinger equations that, in principle, needs to be solved iteratively
requiring prohibitive amounts of computational resources. To avoid this
complication, we truncate these loop contributions by employing in our present
application only the usual (on-shell) $\gamma^\mu$ and $\sigma^{\mu\nu}k_\nu$
pieces for the nucleon current within \textit{all} the loops. This
approximation step thus linearizes the nucleon-current contributions. Second,
we note that the current piece $\tJ^\mu_s$ that appears in the $s$-channel part
of (\ref{eq:MmuWithT}) obeys the same WTI (\ref{eq:NWTI}) as the full current,
i.e.,
\begin{equation}
  k_\mu J^\mu = k_\mu \tJ^\mu_s= S^{-1}(p+k)Q_N -Q_N S^{-1}(p)~,
  \label{eq:sWTI}
\end{equation}
since the their difference in (\ref{eq:JmuDetail}) is purely transverse. The
detailed expressions for $\tJ^\mu_s$, moreover, mainly are comprised of
\textit{longitudinal} contributions that arise from the photon being attached
to the self-energy loop of the nucleon propagator [cf.\
Figs.~\ref{fig:NcurrentC}(b) and \ref{fig:hadrondressing}(a)].

To reproduce the WTI (\ref{eq:sWTI}), it is
convenient to utilize the Ball-Chiu current \cite{BallChiu1980},
\begin{align}
  J^\mu_s &=(2p+k)^\mu \frac{S^{-1}(p+k)Q_N-Q_NS^{-1}(p)}{s^2-p^2}
\nonumber\\
&\mbox{}\quad
+\left[\gamma^\mu -
\frac{(2p+k)^\mu}{s-p^2}\dslash{k}\right]Q_N\frac{A(s)+A(p^2)}{2}~,
 \label{eq:JsDefined}
\end{align}
where $s=(p+k)^2$ and $A$ is one of two scalar dressing functions ($B$ being
the other) resulting from the generic form
\begin{equation}
  S(p) = \frac{1}{\dslash{p} A(p^2)-mB(p^2)}
\end{equation}
of the dressed nucleon propagator; $m$ is the nucleon mass. The dressing
functions $A(p^2)$ and $B(p^2)$ are constrained to produce a unit residue at
the nucleon pole where $p^2=m^2$ \cite{Haberzettl97,AA95}. By construction,
$J^\mu_s$ is the minimal current that satisfies the WTI (\ref{eq:sWTI}) for
fully dressed propagators \textit{and} it is both non-singular and symmetric.
Without lack of generality, therefore, we may write
\begin{equation}
  \tJ_s^\mu = J^\mu_s + \tT^\mu_s~,
  \label{eq:tildeJsSplit}
\end{equation}
where $\tT^\mu$ is the transverse remainder defined by this relation.

Within the $s$-channel context of Eq.~(\ref{eq:MmuWithT}), the Ball-Chiu
current provides a particularly simple expression when considering the
half-on-shell situation, with an incoming nucleon spinor $u(p)$ on the right
and an outgoing propagator $S(p+k)$ on the left. One easily finds
\begin{align}
SJ^\mu_s u &=
\Bigg[\frac{1}{\dslash{p}+\dslash{k}-m}\left(\gamma^\mu + \frac{i\sigma^{\mu\nu}k_\nu}{2m}\kappa_1\right)
\nonumber\\
&\qquad\mbox{}
  + i\sigma^{\mu\nu}k_\nu \frac{\kappa_2-\kappa_1}{s-m^2}\Bigg]Q_N \, u(p)~,
  \label{eq:JsRonshellj}
\end{align}
where $p^2=m^2$. Albeit written in a somewhat different way, this expression is
equivalent to Eq.~(17) of Ref.~\cite{HHN2011}, and the two independent
coefficient functions $\kappa_i=\kappa_i(s)$ ($i=1,2$) that are directly
related to the propagator dressing functions $A$ and $B$ are given in Eq. (19)
of Ref.~\cite{HHN2011}. We omit their details because we will not make use of
them here. We only mention that the on-shell values at $s=m^2$ for both
coefficients are identical, i.e., $\kappa_1(m^2)=\kappa_2(m^2) =A(m^2)-1$, and
  that one easily finds that $(\kappa_2-\kappa_1)/(s-m^2)$ possesses a finite
  limit for $s\to m^2$ if we assume that the dressing functions $A$ and $B$ are
  analytic functions in the vicinity of $s=m^2$.
This means they both vanish in the structureless limit [where $A(m^2)=1$], thus
leaving in (\ref{eq:JsRonshellj}) only the usual $\gamma^\mu$ Dirac current
together with a structureless propagator. All effects of the dressing thus
reside in the terms that depend on the $\kappa_i$ whose overall contributions
are manifestly transverse. In the present application, we will absorb the
$\kappa_i$ ($i=1,2$) dependence in some fit parameters.

We note in this context that the Ball-Chiu current $J^\mu_s$ does not fully
contain the anomalous-moment contribution of the usual Pauli part of the
on-shell nucleon current. These anomalous contributions arise from other
transverse dressing mechanisms. Writing the on-shell matrix element of the full
nucleon current between nucleon spinors in the usual manner as
\begin{equation}
 \bar{u} J^\mu u= e \bar{u} \left(\gamma^\mu \delta_N+\frac{i\sigma^{\mu\nu}k_\nu}{2m}\kappa_N\right) u~,
\end{equation}
where $e$ is the fundamental charge unit, $\kappa_N$ is the anomalous moment of
the nucleon, and $\delta_N=1,0$ for the proton and neutron, respectively, the
anomalous part can be written as
\begin{align}
  e \,\bar{u}\left( \frac{i\sigma^{\mu\nu}k_\nu}{2m}\kappa_N\right)u
  &=\bar{u} \Bigg(e\delta_N\frac{i\sigma^{\mu\nu}k_\nu}{2m}\tkappa_0
  \nonumber\\
  &\qquad\mbox{}
  +\tilde{T}^\mu+\bra{F} G_0 B^\mu_\tra\Bigg)u
  \label{hh7-kappa1}~,
\end{align}
where $\tkappa_0=\kappa_1(m^2)=\kappa_2(m^2)=A(m^2)-1$. This result follows
from Eq.~(\ref{eq:JmuDetail}) utilizing Eq.~(\ref{eq:tildeJsSplit}) and the
expression (\ref{eq:JsRonshellj}) performing expansions of $\kappa_i(s)$ around
$s=m^2$.  In principle, this equation is exact if $\tilde{T}^\mu$ and
$B^\mu_\tra$  could be calculated without approximations. We see here that for
the proton, the Ball-Chiu term $\tkappa_0$ contributes partially to the
anomalous moment, but for the neutron, it does not contribute at all.

In Ref.~\cite{HHN2011}, it was advocated to put $\tilde{T}^\mu=0$
and ensure the anomalous contributions by adding a contact term to the
approximation of $M^\mu_c$ to be discussed in the following subsection in the
context of Eq.~(\ref{eq:Mcapprox}). In the present application, however, we
proceed differently. We expand the full dressing functions $\kappa_i$ appearing
in the Ball-Chiu contribution (\ref{eq:JsRonshellj}) around their on-shell
points and use the resulting coefficients as fit parameters. In addition, we
assume that we can approximate $\tilde{T}^\mu$ by a single transverse contact
term whose operator structure is given by $\sigma^{\mu\nu}k_\nu$ alone. It is
the corresponding coefficient, in particular, that ensures that we can
reproduce the anomalous moments. For the half-on-shell matrix element of
$\tJ_s^\mu$ appearing in the $s$-channel pole term of Eq.~(\ref{eq:MmuWithT}),
we then obtain the \textit{approximation}
\begin{align}
S\tJ^\mu_s u \to S\tJ^\mu_s u &=e\Bigg[\frac{1}{\dslash{p}+\dslash{k}-m}\left(\delta_N\gamma^\mu + \frac{i\sigma^{\mu\nu}k_\nu}{2m}\kappa_0\right)
\nonumber\\
&\qquad\mbox{}
  +\frac{t^{\mu\nu}k_\nu}{4m^2}C_1  + \frac{i\sigma^{\mu\nu}k_\nu}{4m^2}C_2 \Bigg] u(p)~,
  \label{eq:newpar3}
\end{align}
where
\begin{equation}
t^{\mu\nu}k_\nu=\frac{\gamma^\mu (s-m^2)-(2p+k)^\mu \dslash{k}}{2m}
\end{equation}
is a transverse current operator that follows from the expansions of the
Ball-Chiu contributions. For each nucleon, this approximation contains three
dimensionless coefficients: a real parameter $\kappa_0$, and two complex
numbers $C_1$ and $C_2$ for the contact contributions.

We emphasize that the approximation just discussed \textit{only}
pertains to the true nucleon current. In the full coupled-channels treatment,
the structure of Eq.~(\ref{eq:MmuWithT}) also comprises contributions from
currents $\tJ^\mu_s$ that describe nucleon-resonance transitions. Such
transition currents, however, must necessarily be transverse and thus
correspond only to the first two diagrams on the right-hand side of
Fig.~\ref{fig:NcurrentC}(b). In the present applications, thus, we describe
these transition currents by the point vertices of the nucleon-resonance
photo-transition Lagrangians given in Eqs.~(\ref{1hfNr}) and (\ref{3hfNr}) of
the Appendix.

We also emphasize that the present approximate treatment of the
$s$-channel term (\ref{eq:newpar3}) that permits one to describe the
corresponding fully dressed contribution in terms of three parameters requires
the complete reciprocal consistency between the nucleon current and the
photoproduction current, as derived in Ref.~\cite{HHN2011}. The approximation scheme of
Ref.~\cite{Haberzettl06}, by contrast, is ambiguous in its treatment of
certain transverse current contributions that have no bearing on gauge
invariance. Specifically, the undetermined transverse current $T^\mu$
appearing in Eq.~(21) of Ref.~\cite{Haberzettl06} was taken to be zero in the
preliminary applications reported there. To achieve complete structural
equivalence with the present results, one finds that $T^\mu$ must be chosen as
$T^\mu = UG_0 (M^\mu_c)_\tra$ instead if one wants to preserve the equivalence
even when making approximations.

\subsubsection{Approximating the four-point contact current $M^\mu_c$}

In view of the fact that the FSI loop in (\ref{eq:MmuWithT}) is transverse and
that the approximation of the $s$-channel current $\tJ^\mu_s$ by the Ball-Chiu
current $J^\mu_s$ leaves the corresponding WTI unchanged, the entire burden for
reproducing the generalized WTI for the production current $M^\mu$ rests with
the contact-type four point current $M^\mu_c$ whose four-divergence, by
construction, must be the same as that of the interaction current $M^\mu_\IC$
in Eq.~(\ref{eq:gWTIint}), i.e.,
\begin{equation}
  k_\mu\,M^\mu_\IC =k_\mu M^\mu_c= -\tF_s e_i + \tF_u e_f + \tF_t e_\pi~.
  \label{eq:gWTIc}
\end{equation}
As far as preserving gauge invariance is concerned, therefore, the situation is
very much the same as in the approach of Ref.~\cite{Haberzettl06}, even though
the details of the present formulation are somewhat different. In other words,
any approximation of $M^\mu_c$ must satisfy the constraint (\ref{eq:gWTIc}).

The internal dynamical mechanisms of $M^\mu_c$ as depicted in
Fig.~\ref{fig:MwT}(c) are quite involved and cannot, at present, be
incorporated fully in numerical applications. In Ref.~\cite{Haberzettl06}, it
was discussed in detail how to find approximations of the mechanisms within
$M^\mu_c$ at any desired level of sophistication while still reproducing
(\ref{eq:gWTIc}). For the present application, we simply quote the results of
\cite{Haberzettl06} valid when the $\pi NN$ vertices are described by
phenomenological form factors, as it is the case here.

Phenomenologically, the $\pi NN$ vertex stripped of its isospin dependence can be written as
\begin{equation}
\tF_x = \gpi  \frac{\gamma_5\dslash{q}}{2m} f_x~,
 \label{eq:tFdef}
\end{equation}
where we take here pure pseudovector coupling, with $\gpi$ being the physical
coupling constant. The subscript $x=s$, $u$, or $t$ indicates the kinematic
context; $f_x$ is the phenomenological form factor for the corresponding
reaction channel normalized to unity when all hadron legs are on-shell and $q$
is the four-momentum of the outgoing meson.  Following
Ref.~\cite{Haberzettl06}, we define an auxiliary current
\begin{align}
C^\mu &=  - \, e_\pi \frac{f_t-\hat{F}}{t-q^2}  (2q-k)^\mu - e_f \frac{f_u-\hat{F}}{u-p'^2}(2p'-k)^\mu
\nonumber \\
 & \qquad\mbox{} - \, e_i  \frac{f_s-\hat{F}}{s-p^2} (2p+k)^\mu~,
\end{align}
with
\begin{equation}
\hat{F}= 1-\hat{h}\left(1-\delta_s f_s\right) \left(1-\delta_u
f_u\right)\left(1-\delta_t f_t\right)~,
\label{eq:Fhat}
\end{equation}
where the four-momenta are those given in Fig.~\ref{fig:gammaN}. The constant
$\delta_x$ is unity if the corresponding $x$-channel contributes to the
reaction in question, and zero otherwise. The parameter $\hat{h}$ may be an
arbitrary (complex) function, $\hat{h}=\hat{h}(s,u,t)$, possibly subject to
crossing-symmetry constraints. In the present work, it is simply taken as a fit
constant for the sake of simplicity. The four-divergence of the auxiliary
current evaluates to
\begin{equation}
k_\mu C^\mu =  - f_s e_i  + f_u e_f +f_t e_\pi~.
\label{eq:k.C}
\end{equation}
We emphasize that $C^\mu$ is \textit{non-singular} since, by construction, the
propagator singularities are canceled by the corresponding zeros of
$f_x-\hat{F}$. In other words, $C^\mu$ is a true contact current.

The gauge-invariance preserving (GIP) approximation of the contact
current $M^\mu_c$ then may be chosen as \cite{Haberzettl06,HHN2011}
\begin{align}
M^\mu_c \to M^\mu_c  &=  \gpi \frac{\gamma_5(\dslash{q}-\beta \dslash{k})}{2m}\, C^\mu \nonumber\\
&\mbox{}\qquad -  \gpi \frac{\gamma_5 \gamma^\mu}{2m}  \left(e_\pi f_t - \beta k_\rho C^\rho\right)~,
\label{eq:Mcapprox}
\end{align}
where $\beta$ is a free fit parameter. One can easily check that this
choice for $M^\mu_c$ satisfies the gauge-invariance condition (\ref{eq:gWTIc}).
Essential in this respect is the relation (\ref{eq:k.C}). Note that the term
proportional to the pion charge $e_\pi$ in (\ref{eq:Mcapprox}) is a dressed
version of the familiar Kroll-Ruderman contact current where the dressing form
factor $f_t$ is that of the $t$-channel amplitude $M^\mu_t$. The original
(undressed) Kroll-Ruderman term survives in this GIP current since $f_t =
1+(f_t-1)$, i.e., the dressing is given by the additional $f_t-1$ contribution.

We emphasize that the present formulation is equally valid for real and virtual
photons. For pion photoproduction with real photons, in particular, only
transverse currents contribute to the physical amplitude. The longitudinal loop
contributions shown for $M^\mu_c$ in Fig.~\ref{fig:MwT}(c), therefore, are
projected out and only the (bare) Kroll-Ruderman current and the loop over the
the five-point interaction current $U^\mu$ of Fig.~\ref{fig:MwT}(d) remain.
Hence, subtracting the usual (bare) Kroll-Ruderman term from the GIP current
(\ref{eq:Mcapprox}), the remaining expression is a phenomenological
approximation of the latter loop contribution for real photons.

Finally, we mention that when we consider intermediate channels other than $\pi
N$ (cf.\ Sec.~\ref{sec:furtherdetail} below), the corresponding contact currents
$M^\mu_c$ that appear in the transverse loops $TG_0 B^\mu_\tra$ of
Eq.~(\ref{eq:MmuWithT}) are easily constructed in analogy to the explicit $\pi
N$ case given here by using the generic constraint (\ref{eq:gWTIc}).

\subsubsection{Covariant three-dimensional reduction}

The present formalism is a fully covariant approach. To calculate any reaction
amplitude in a full four-dimensional framework is a daunting numerical task.
Therefore, in the present work, we approximate the full meson-baryon two-body
propagator $G_0$ by the Kadyshevsky propagator \cite{BbS}, as specified in
Eqs.~(\ref{eq:BbS}) and (\ref{eq:GBbS}) in the next section. This propagator,
together with an implied energy delta function, restricts the propagation of
the intermediate meson and baryon to their respective energy shells, thus
reducing the four-dimensional loop integration to a three-dimensional one
without destroying the covariance of the equation. This reduction is also
necessary to be able to match up the hadronic J\"ulich-model input with the
present formalism.

\section{\boldmath Making the J\"ulich $\pi N$ model compatible with the present approach}\label{sec:Juelich}

In the present work, we employ the J\"ulich dynamical coupled-channels $\pi N$
model \cite{Schutz98,Krehl00,Gasparyan03} to account for the hadron dynamics.
As mentioned in the Introduction, this model is formulated within time-ordered
perturbation theory (TOPT) which is a non-covariant three-dimensional
formalism \cite{Schweber}.  We, therefore, need to provide a procedure to match
it to the covariance of the present photoproduction formalism.

To this end, we first note that the J\"ulich $T$-matrix satisfies the
three-dimensional Lippmann-Schwinger-type equation
\begin{align}
T_{\TO}(\bv{p}', \bv{p}; z)  &=  V_{\TO}(\bv{p}', \bv{p}; z)
\nonumber\\
&\mbox{}\hspace{-18mm}+
\int d^3p'' \, V_{\TO}(\bv{p}', \bv{p}''; z) \, G_{\TO}(\bv{p}'', z)
T_{\TO}(\bv{p}'', \bv{p}; z)~, \label{eq:TO}
\end{align}
where $z$ denotes the total-energy variable and TO indicates that all entities
result from the time-ordered formalism. The intermediate pion-nucleon
propagator reads
\begin{equation}
G_{\TO}(\bv{p}'', z) = \frac{1}{z - E(\bv{p}'') - \omega(\bv{p}'') + i0}~,
\label{eq:GTO}
\end{equation}
where $E(\bv{p}'') \equiv \sqrt{m^2 + \bv{p}''^2}$ stands for the energy of the
nucleon and $\omega(\bv{p}'') \equiv \sqrt{m_\pi^2 + \bv{p}''^2}$ for the
energy of the pion, with the mass $m_\pi$. The complex notation $i0$ indicates
the physical limit to the upper edge of the scattering cut to provide proper
boundary conditions. Equation (\ref{eq:TO}) written in the c.m.\ system is
structurally very similar to the Bethe-Salpeter equation (\ref{eq:TBSeq}) after
it has been subjected to a covariant three-dimensional reduction. Therefore, to
match the two, one needs to ensure that Eq.~(\ref{eq:TO}) transforms
covariantly away from the c.m.\ system.

\begin{table}[t!]
\caption{\label{tab:para_cut} Cutoff parameters  (in MeV) in the form factors
of Eqs.~(\ref{eq:ff}) and (\ref{eq:secondff}).}
\renewcommand{\arraystretch}{1.3}
\begin{tabular*}{\columnwidth}{@{\extracolsep\fill}cccc}
\hline\hline
 $\Lambda$ & $\Lambda_B$ & $\Lambda_\pi$ & $\Lambda_v$  \\ \hline
 $600$ & $725$ & $1343$ & $1645$ \\
\hline\hline
\end{tabular*}
\end{table}

\begin{table}[t!]
\caption{\label{tab:para_res_1hf} Effective electromagnetic coupling constants
for isospin $1/2$ resonances. The subscripts $p$ and $n$ denote the resonance
coupling to $\gamma p$ and $\gamma n$, respectively.}
\renewcommand{\arraystretch}{1.3}
\begin{tabular*}{\columnwidth}{@{\extracolsep\fill}crrrr}
\hline\hline
              & $D_{13}(1520)$ & $S_{11}(1535)$ & $S_{11}(1650)$ & $P_{13}(1720)$ \\ \hline
 $~~g_p^{(1)}$  & $-1.23~~~$     & $-1.14~~~$     & $0.02~~~$     & $  4.23~~~$  \\
 $~~g_p^{(2)}$  & $1.58~~~$      &                &                & $-12.56~~~$  \\
 $~~g_n^{(1)}$  & $2.69~~~$      & $ 0.21~~~$     & $-0.17~~~$     & $ -0.73~~~$  \\
 $~~g_n^{(2)}$  & $-3.58~~~$     &                &                & $  2.52~~~$  \\
\hline\hline
\end{tabular*}
\end{table}

\begin{table}[t!]
\caption{\label{tab:para_res_3hf} Effective electromagnetic coupling constants
for isospin $3/2$ resonances.}
\renewcommand{\arraystretch}{1.3}
\begin{tabular*}{\columnwidth}{@{\extracolsep\fill}crrrr}
\hline\hline
            & $P_{33}(1232)$ & $S_{31}(1620)$ & $D_{33}(1700)$ & $P_{31}(1910)$ \\ \hline
 $~~g^{(1)}$  & $2.01~~~$      & $1.63~~~$      & $-6.56~~~$     & $-1.44~~~$  \\
 $~~g^{(2)}$  & $4.48~~~$      &                & $5.38~~~$      &   \\
\hline\hline
\end{tabular*}
\end{table}

We define now
\begin{align}
V(\bv{p}',\bv{p};\sqrt{s}) &\equiv (2\pi)^3 \rho(\bv{p}')\,  V_{\TO}(\bv{p}', \bv{p}; z)\, \rho(\bv{p})~,
\label{eq:MRF1}
\\
T(\bv{p}', \bv{p}; \sqrt{s}) &\equiv (2\pi)^3 \rho(\bv{p}')\, T_{\TO}(\bv{p}', \bv{p}; z)\, \rho(\bv{p})~,
 \label{eq:MRF2}
\end{align}
where
\begin{equation}
  \rho(\bv{p}) =  2\sqrt{E(\bv{p})\omega(\bv{p}) }
\end{equation}
corrects the non-covariant normalizations of the TOPT plane-wave states and
where the invariant mass $\sqrt{s}$ is defined as $\sqrt{s}\equiv z$ in the
c.m.\ system. Equation (\ref{eq:TO}) then may be recast in the equivalent form
\begin{align}
T(\bv{p}', \bv{p}; \sqrt{s})  &=  V(\bv{p}', \bv{p}; \sqrt{s})
\nonumber\\
&\mbox{}\hspace{-18mm}
+ \int \frac{d^3p''}{(2\pi)^3} \, V(\bv{p}', \bv{p}''; \sqrt{s}) \,
G_0(\bv{p}'', \sqrt{s}) \, T(\bv{p}'', \bv{p}; \sqrt{s})~, \label{eq:BbS}
\end{align}
where
\begin{align}
G_0(\bv{p}'', \sqrt{s})  \equiv  \frac{1}{\rho^2(\bv{p}'')}\;\frac{1}{\sqrt{s} - E(\bv{p}'')
-\omega(\bv{p}'') + i0}~. \label{eq:GBbS}
\end{align}
Equation (\ref{eq:BbS}) is formally identical to the Kadyshevsky equation
\cite{BbS} which is the result of a particular \textit{covariant}
three-dimensional reduction of the Bethe-Salpeter equation that preserves
elastic unitarity. Hence, utilizing the time-ordered J\"ulich $T$-matrix in the
redefined form (\ref{eq:MRF2})  is completely consistent with the covariant
three-dimensional reduction of the photoproduction current given in
Eq.~(\ref{eq:MmuWithT}). For the corresponding reduced meson-baryon propagator
one must use the form (\ref{eq:GBbS}), of course.

\begin{table}[t!]
\caption{\label{tab:para_CCk} Dimensionless parameters $C_1$, $C_2$, and
$\kappa_0$ in the electromagnetic nucleon current, Eq.~(\ref{eq:newpar3}). }
\renewcommand{\arraystretch}{1.3}
\begin{tabular*}{\columnwidth}{@{\extracolsep\fill}lccc}
\hline\hline
& $C_1$ & $C_2$ & $\kappa_0$\\ \hline
proton & $2.07\,{e}^{-1.74 i}$  & $34.62\,{e}^{0.09 i}$  & $-4.24~~$ \\
neutron & $12.35\,{e}^{2.52 i}$ & $25.96\,{e}^{2.79 i}$  & $3.24$ \\
\hline\hline
\end{tabular*}
\end{table}

\begin{table}[t!]
\caption{\label{tab:para_hat} Dimensionless parameters $\hat{h}$ and $\beta$ in
the contact current $M^\mu_c$, Eq.~(\ref{eq:Mcapprox}), for the various
meson-baryon channels. }
\renewcommand{\arraystretch}{1.3}
\begin{tabular*}{\columnwidth}{@{\extracolsep\fill}crrrrr}
\hline\hline
            & $\pi^+n$ & $\pi^0p~$ & $\pi^-p$ & $\eta p~$ & $\pi \Delta~$ \\ \hline
$~~\hat{h}$ & $1.43$   & $1.44$    & $2.01$   & $-10.04$  & $0.74$ \\
$~~\beta$   & $-0.17$  & $-1.43$   & $-0.10$  & $3.45$    & $-1.75$ \\
\hline\hline
\end{tabular*}
\end{table}

We recall in this context that pions and nucleons are used here as simplified
generic tags for all mesons and baryons incorporated in the J\"ulich model.
Therefore, all entities discussed here need to be considered as elements of
matrices labeled by meson-baryon channels and all equations contain implied
summations over all elements of this channel space. Moreover, for
quasi-two-body channels $\sigma N$, $\rho N$, and $\pi\Delta$, the propagator
(\ref{eq:GBbS}) is to be slightly modified by including the self-energy of the
corresponding quasi-free particles \cite{Schutz98,Krehl00}.

%=======================================================
\begin{figure}[t!]\centering
  \includegraphics[width=.8\columnwidth,clip=]{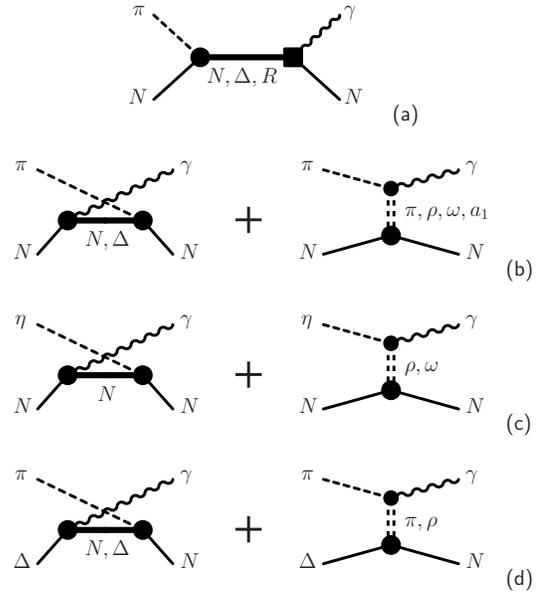}
  \caption{\label{fig:channels}%
  Hadronic states taken into account in the present model (see
  Sec.~\ref{sec:furtherdetail} for details): (a) intermediate $s$-channel
  baryons  for $\gamma N \to \pi N$ ($R$ subsumes all resonances except the
  $\Delta$) contributing to Fig.~\ref{fig:MwT}(a); $u$- and $t$-channel
  contributions as they appear in $B^\mu$ of Fig.~\ref{fig:MwT}(b) for
  (b) $\gamma N \to \pi N$, (c) $\gamma N \to \eta N$, and (d) $\gamma N \to
  \pi \Delta$. Diagrams (c) and (d) only contribute within the FSI loop. }
\end{figure}
%=======================================================

\begin{figure*}[tb] \centering
\includegraphics[width=0.7\textwidth,clip=]{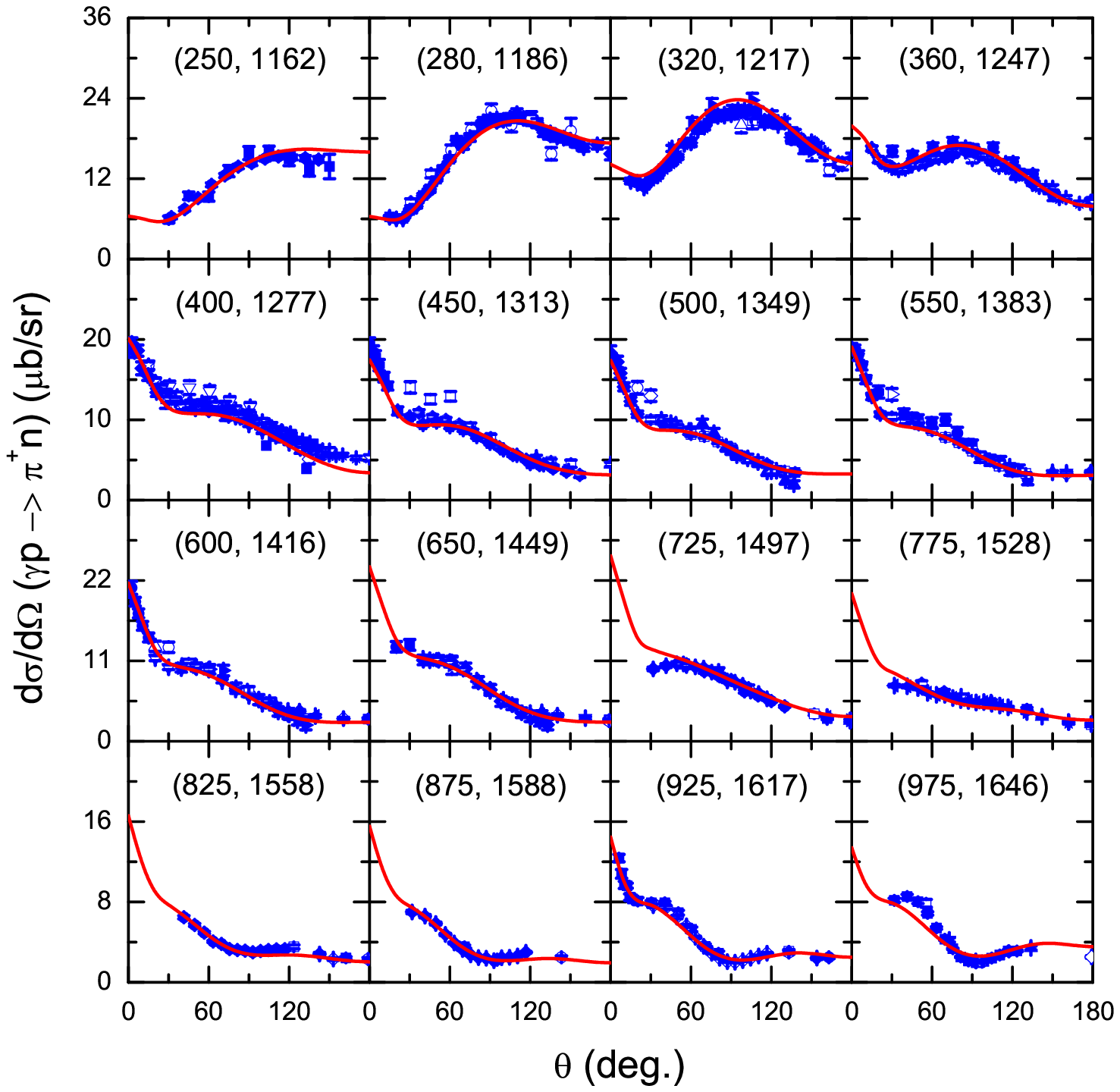}
 \caption{\label{fig:dsig_pi+n} (Color online) Differential cross section for
$\gamma p \to \pi^+ n$ as a function of the scattering angle. The first and
second number in each pair of parentheses correspond to the photon laboratory
momentum and the $\pi N$ c.m.\ energy, respectively. Data are taken from
Refs.~\cite{Dugger09,Ahrens04,SAID}.}
\end{figure*}

\begin{figure*}[tb]\centering
\includegraphics[width=0.7\textwidth,clip=]{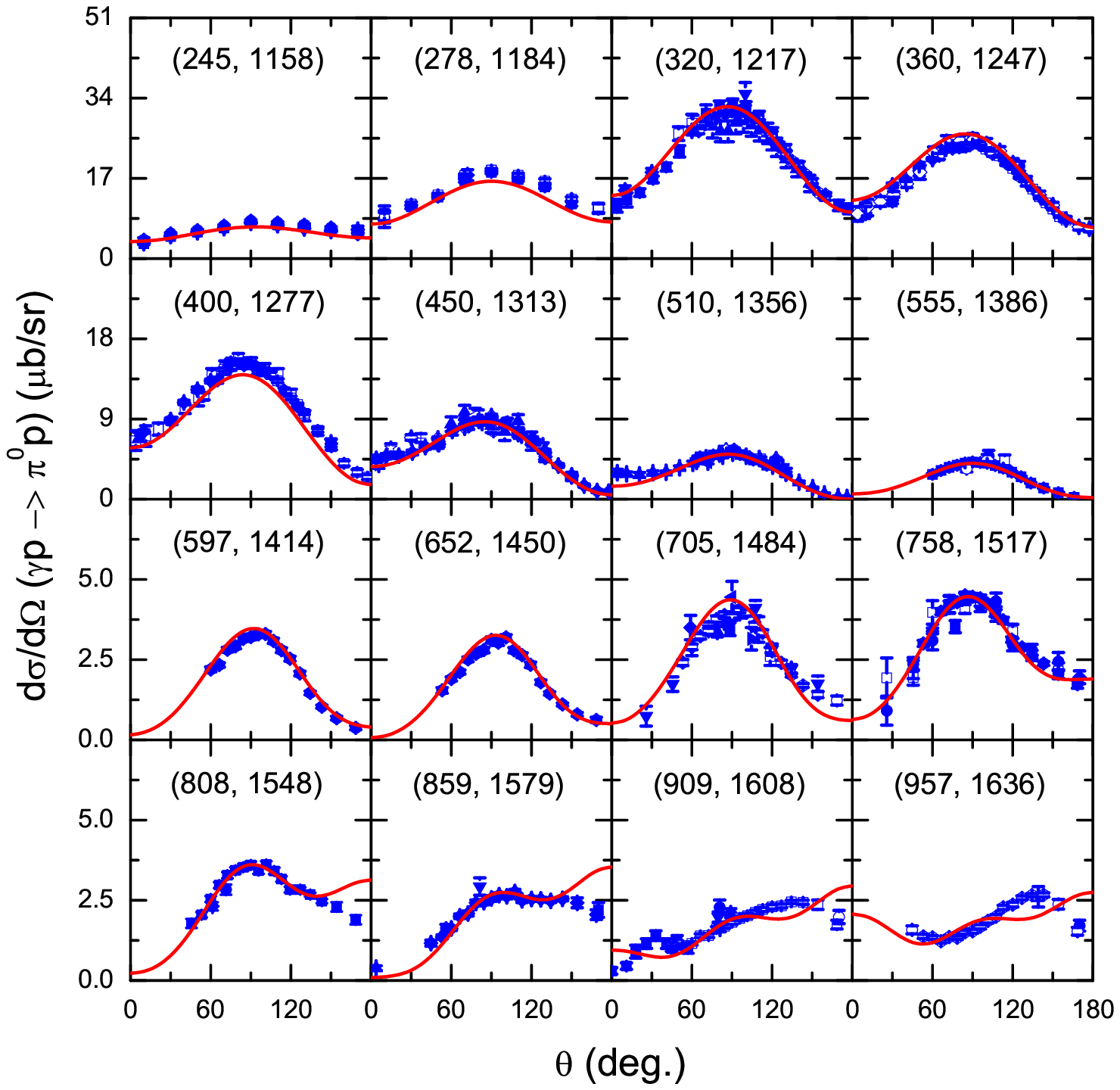}
\caption{\label{fig:dsig_pi0p} (Color online) Differential cross section for
$\gamma p \to \pi^0 p$ as a function of the scattering angle. The first and
second number in each pair of parentheses correspond to the photon laboratory
momentum and the $\pi N$ c.m.\ energy, respectively. Data are taken from
Refs.~\cite{Bartalini05,Bartholomy05,SAID}.}
\end{figure*}

\begin{figure*}[b!] \centering
\includegraphics[width=0.7\textwidth,clip=]{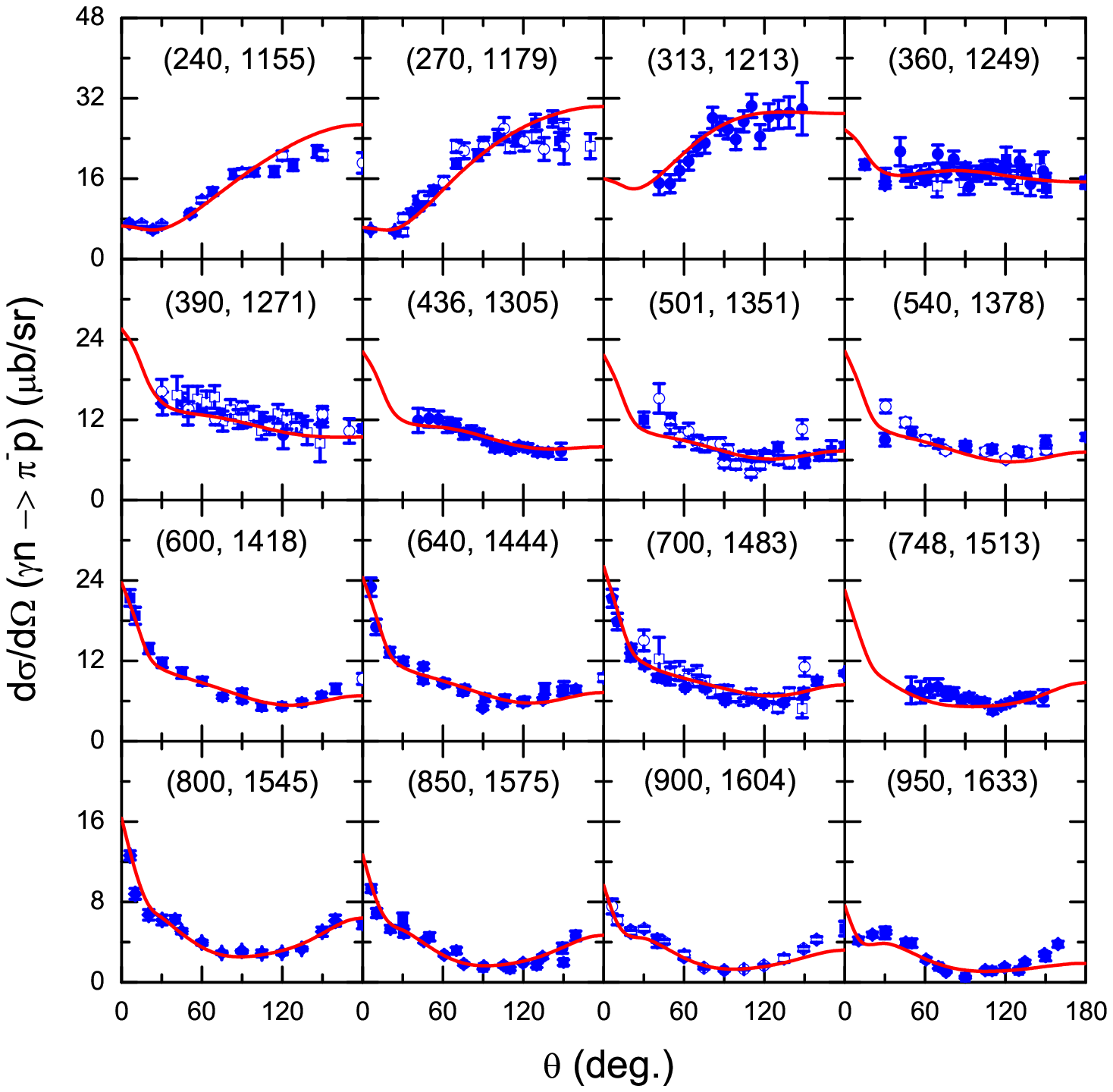}
\caption{\label{fig:dsig_pi-p} (Color online) Differential cross section for
$\gamma n \to \pi^- p$ as a function of the scattering angle. The first and
second number in each pair of parentheses correspond to the photon laboratory
momentum and the $\pi N$ c.m.\ energy, respectively. Data are taken from
Refs.~\cite{Shafi04,SAID}.}
\end{figure*}

\begin{figure*}[tb]\centering
\includegraphics[width=0.7\textwidth,clip=]{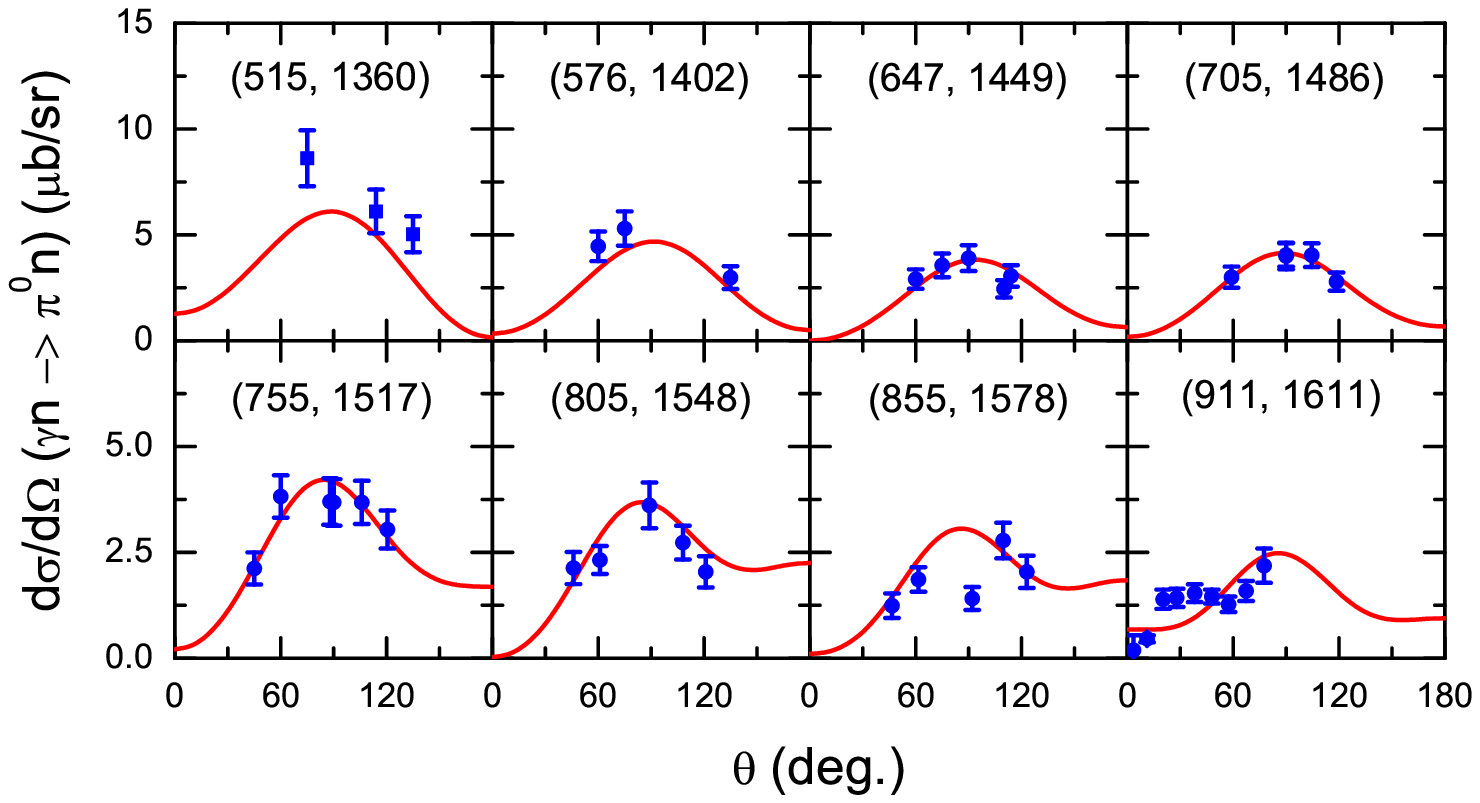}
 \caption{\label{fig:dsig_pi0n} (Color online) Differential cross section for  $\gamma n
\to \pi^0 n$ as a function of the scattering angle. The first and second number
in each pair of parentheses correspond to the photon laboratory momentum and
the $\pi N$ c.m.\ energy, respectively. Data are taken from
Refs.~\cite{Bacci72,Hemmi73}. }
\end{figure*}

\begin{figure*}[tb]\centering
\includegraphics[width=0.7\textwidth,clip=]{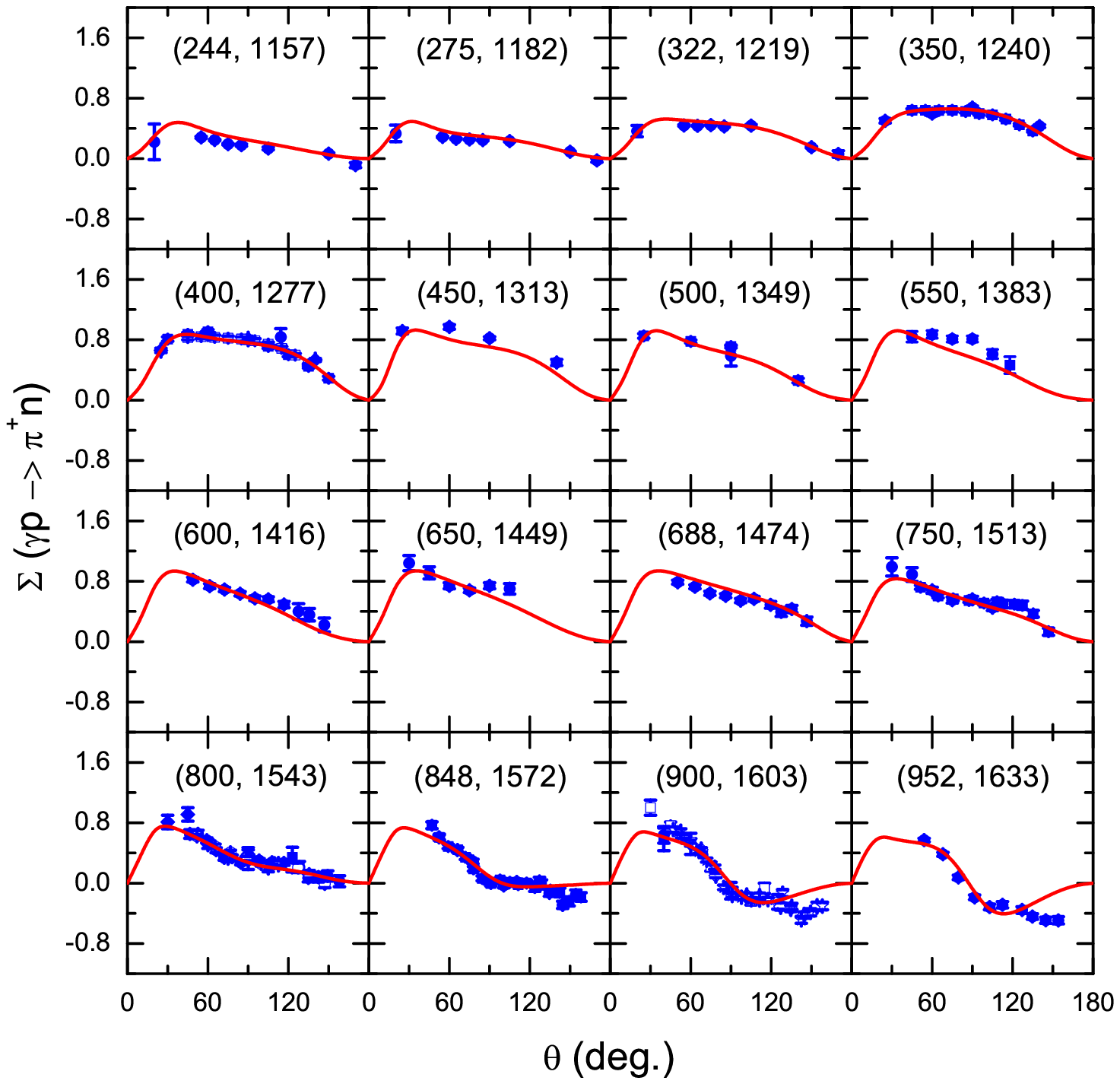}
\caption{\label{fig:sa_pi+n} (Color online) Photon spin asymmetry for $\gamma p
\to \pi^+ n$ as a function of the scattering angle. The first and second number
in each pair of parentheses correspond to the photon laboratory momentum and
the $\pi N$ c.m.\ energy, respectively. Data are taken from Ref.~\cite{SAID}.}
\end{figure*}

\begin{figure*}[tb]\centering
\includegraphics[width=0.7\textwidth,clip=]{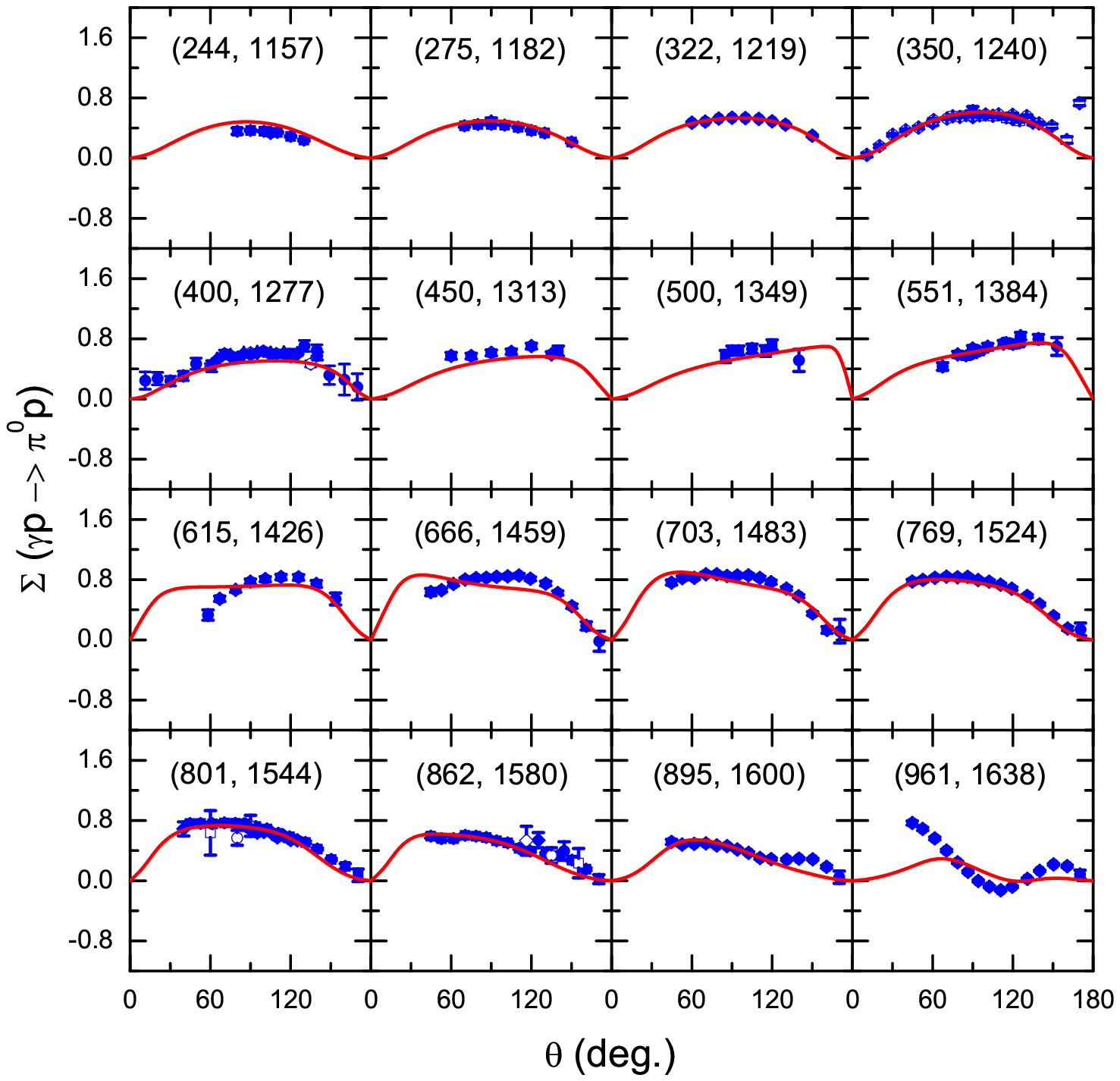}
\caption{\label{fig:sa_pi0p} (Color online) Photon spin asymmetry for $\gamma p
\to \pi^0 p$ as a function of the scattering angle. The first and second number
in each pair of parentheses correspond to the photon laboratory momentum and
the $\pi N$ c.m.\ energy, respectively. Data are taken from
Refs.~\cite{Bartalini05,Elsner09,SAID}.}
\end{figure*}

\begin{figure*}[b!]\centering
\includegraphics[width=0.7\textwidth,clip=]{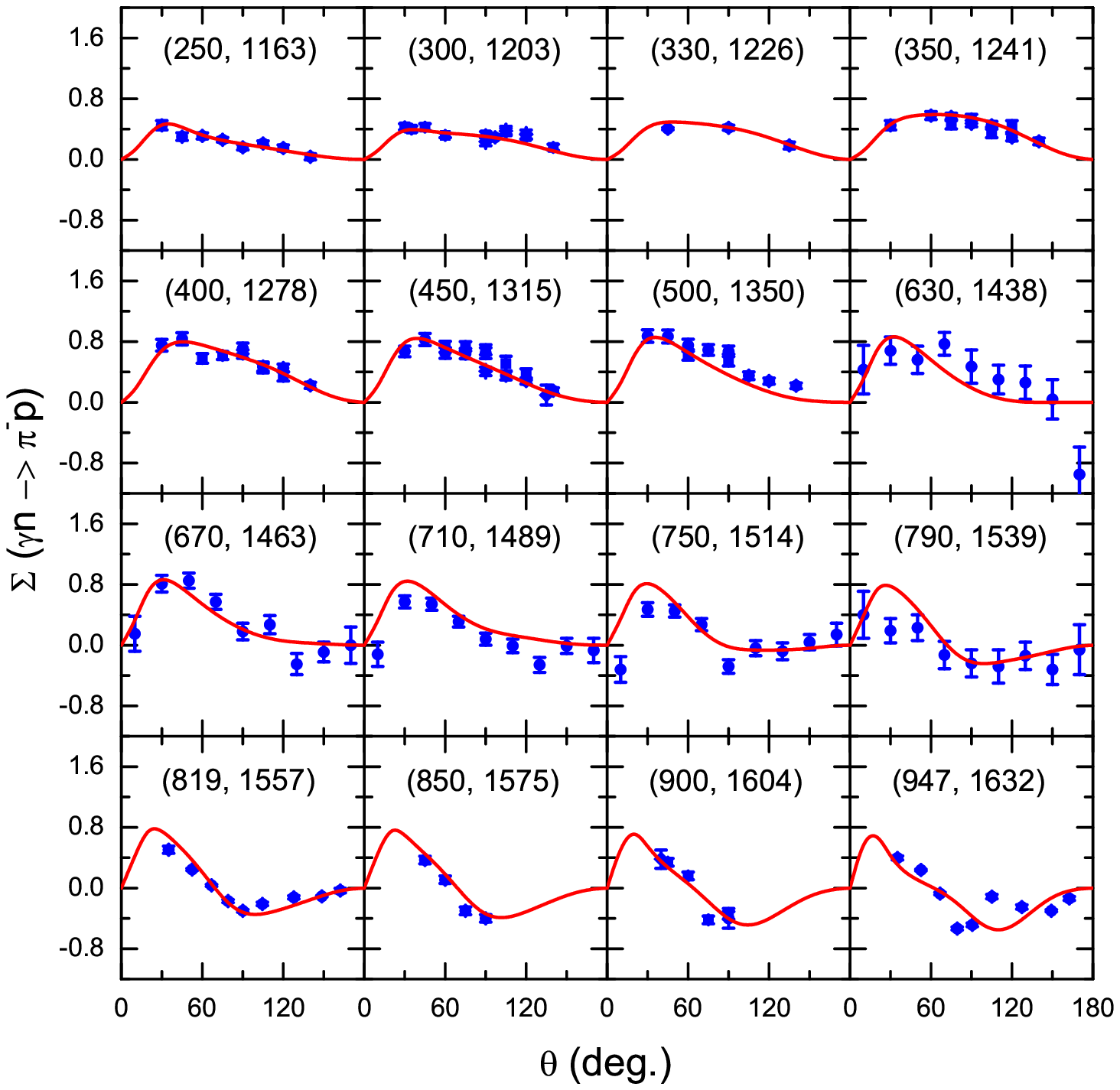}
\caption{\label{fig:sa_pi-p} (Color online) Photon spin asymmetry for $\gamma n
\to \pi^- p$ as a function of the scattering angle. The first and second number
in each pair of parentheses correspond to the photon laboratory momentum and
the $\pi N$ c.m.\ energy, respectively. Data are taken from Ref.~\cite{SAID}.}
\end{figure*}

\begin{figure*}[tb]\centering
\includegraphics[width=0.7\textwidth,clip=]{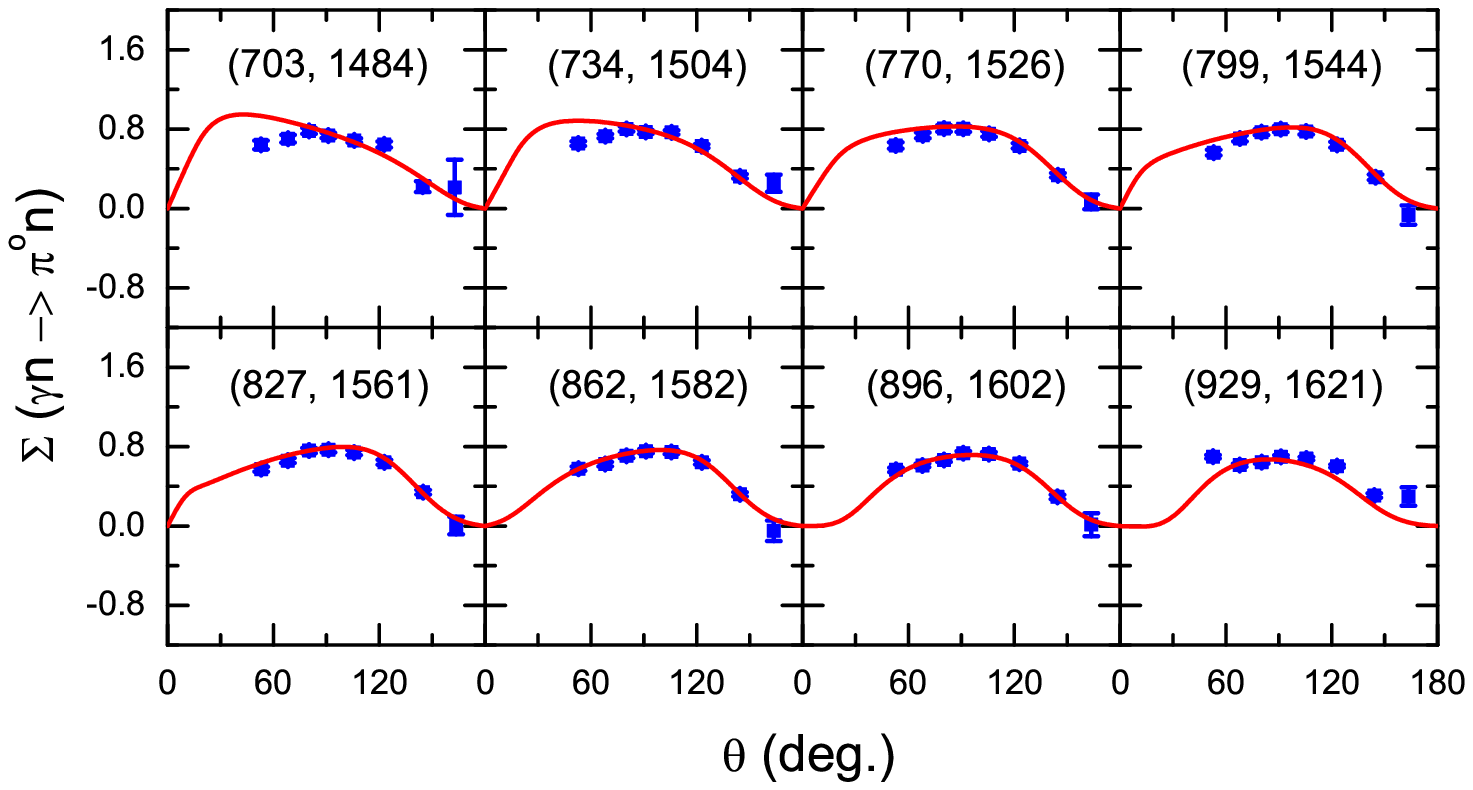}
\caption{\label{fig:sa_pi0n} (Color online) Photon spin asymmetry for
(quasi-free) $\gamma n \to \pi^0 n$ as a function of the scattering angle. The
first and second number in each pair of parentheses correspond to the photon
laboratory momentum and the $\pi N$ c.m.\ energy, respectively. Data are taken
from Ref.~\cite{Salvo09}. }
\end{figure*}

\begin{figure*}[t!]\centering
\includegraphics[width=0.7\textwidth,clip=]{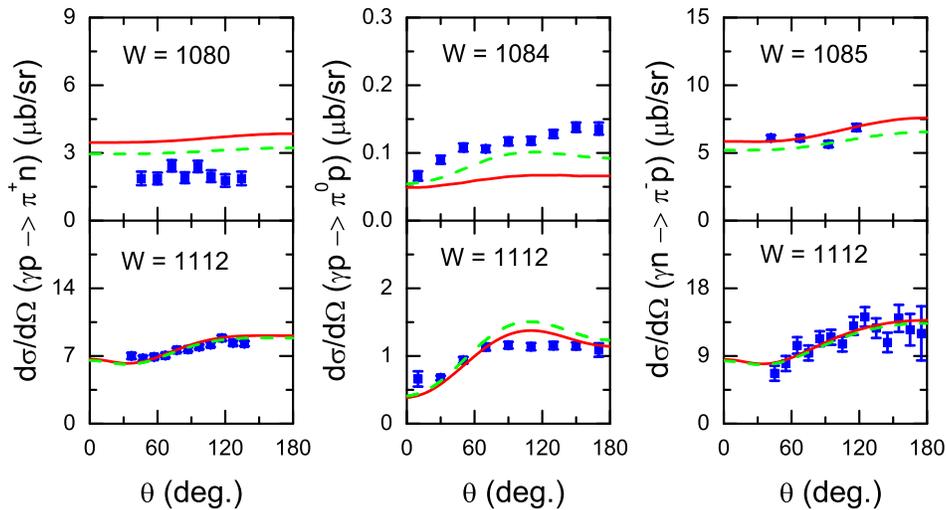}
\caption{\label{fig:dsdo_th} (Color online) Differential cross section for
$\gamma p \to \pi^+ n$, $\gamma p\to \pi^0 p$ and $\gamma n \to \pi^- p$ as a
function of the scattering angle. The c.m.\ energy $W$ is in MeV. The solid
curves show the low-energy results from our model. The dashed curves show
the results from our calculation where the photoproduction kernel is
calculated in particle basis (see text for details). Data are taken from
Refs.~\cite{Korkmaz99,Ahrens04,Schmidt01,Fuchs96,SAID}; they were not included
in the global fit.}
\end{figure*}

\begin{figure*}[t!]\centering
\includegraphics[width=.7\textwidth,clip=]{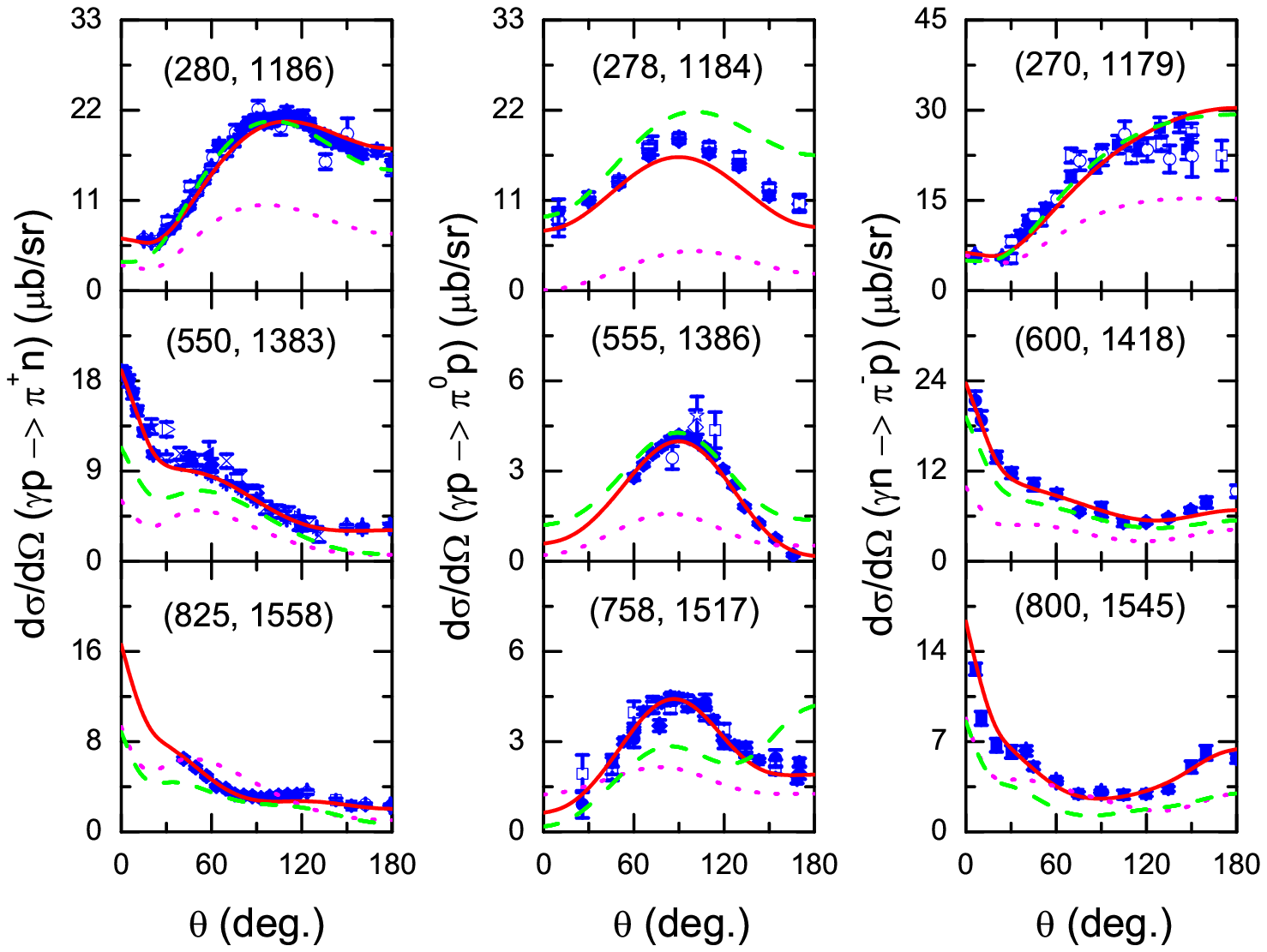}
\caption{\label{fig:dsdo} (Color online) Differential cross section for $\gamma
p \to \pi^+ n$, $\gamma p\to \pi^0 p$ and $\gamma n \to \pi^- p$ as a function
of the scattering angle. The first and second number in each pair of
parentheses correspond to the photon laboratory momentum and the $\pi N$ c.m.\
energy, respectively. The solid curves show the results from the full
calculation. The dotted curves are obtained by switching off the loop integral
[the term proportional to the hadronic amplitude $T$ in
Eq.~(\ref{eq:MmuWithT})]. The dashed curves are obtained by switching off the
contact current other than the Kroll-Ruderman term (the term proportional to
the meson charge $e_\pi$) in $M^\mu_c$ [cf. Eq.~(\ref{eq:Mcapprox})]. Data are
taken from
Refs.~\cite{Dugger09,Ahrens04,Bartalini05,Bartholomy05,Shafi04,SAID}.}
\end{figure*}

\begin{figure*}[tb]\centering
\includegraphics[width=.7\textwidth,clip=]{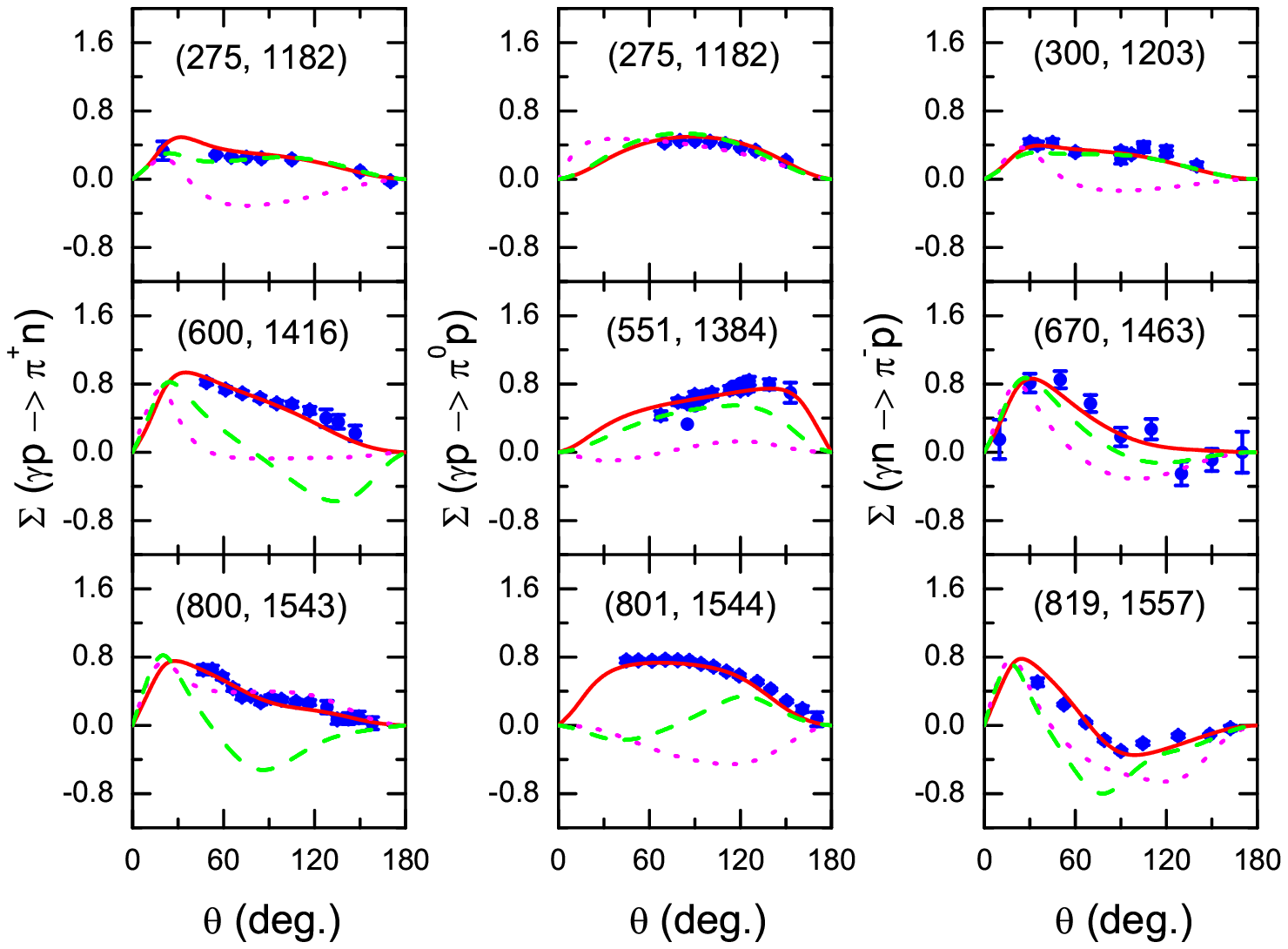}
\caption{\label{fig:sa} (Color online) Photon spin asymmetry for $\gamma p \to
\pi^+ n$, $\gamma p\to \pi^0 p$ and $\gamma n \to \pi^- p$ as a function of the
scattering angle. The first and second number in each pair of parentheses
correspond to the photon laboratory momentum and the $\pi N$ c.m.\ energy,
respectively. The solid curves show the results from the full calculation. The
dotted curves are obtained by switching off the loop integral [the term
proportional to the hadronic amplitude $T$ in Eq.~(\ref{eq:MmuWithT})]. The
dashed curves are obtained by switching off the contact current other than the
Kroll-Ruderman term [the term proportional to the meson charge $e_\pi$) in
$M^\mu_c$ (cf. Eq.~(\ref{eq:Mcapprox})]. Data are taken from
Refs.~\cite{Bartalini05,Elsner09,SAID}.}
\end{figure*}

\begin{figure}[tb]\centering
\includegraphics[width=0.68\columnwidth,clip=]{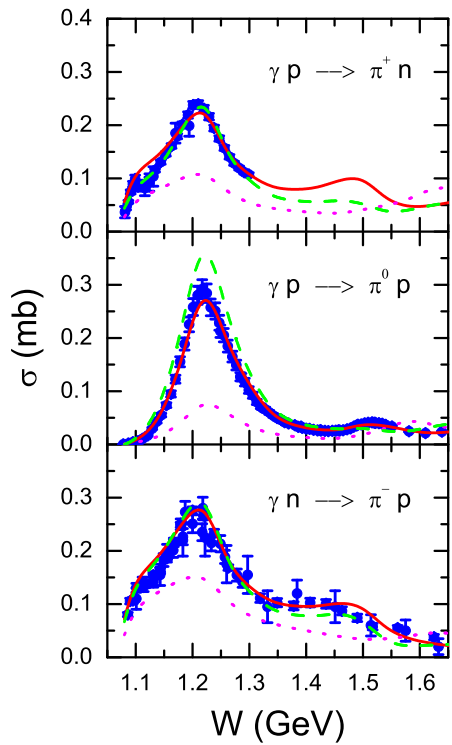}
\caption{\label{fig:sig} (Color online) Total cross section as a function of
the $\pi N$ c.m.\ energy for $\gamma p \to \pi^+ n$, $\gamma p\to
\pi^0 p$ and $\gamma n \to \pi^- p$. The solid curves show the results from the
full calculation. The dotted curves are obtained by switching off the loop
integral in Eq.~(\ref{eq:MmuWithT}). The dashed curves are obtained by switching
off the contact current apart from the Kroll-Ruderman term [the term
proportional to the meson charge $e_\pi$) in $M^\mu_c$ (cf. Eq.~(\ref{eq:Mcapprox})].
Data are taken from Ref.~\cite{SAID} but not included in the fit.}
\end{figure}

\begin{figure*}[t!]\centering
\includegraphics[width=.7\textwidth,clip=]{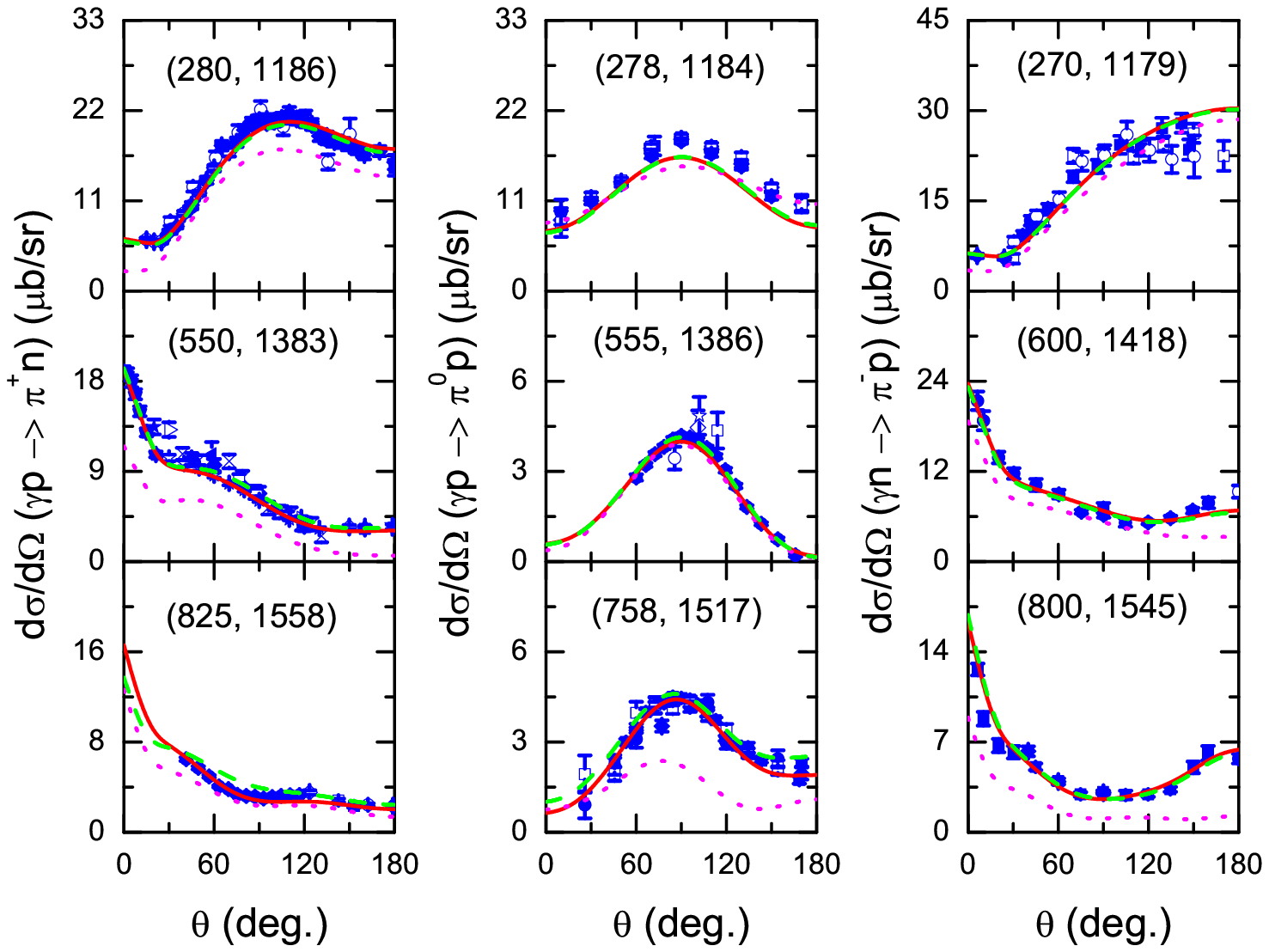}
\caption{\label{fig:dsdo_cc} (Color online) Differential cross section for
$\gamma p \to \pi^+ n$, $\gamma p\to \pi^0 p$ and $\gamma n \to \pi^- p$ as a
function of the scattering angle. The first and second number in each pair of
parentheses correspond to the photon laboratory momentum and the $\pi N$ c.m.\
energy, respectively. The solid curves show the results from the full
calculation. The dotted curves and the dashed curves are obtained by
respectively switching off the $\pi\Delta$ loop integral and the $\eta N$ loop
integral in Eq.~(\ref{eq:MmuWithT}). Data are taken from
Refs.~\cite{Dugger09,Ahrens04,Bartalini05,Bartholomy05,Shafi04,SAID}.}
\end{figure*}

\begin{figure*}[tb]\centering
\includegraphics[width=.7\textwidth,clip=]{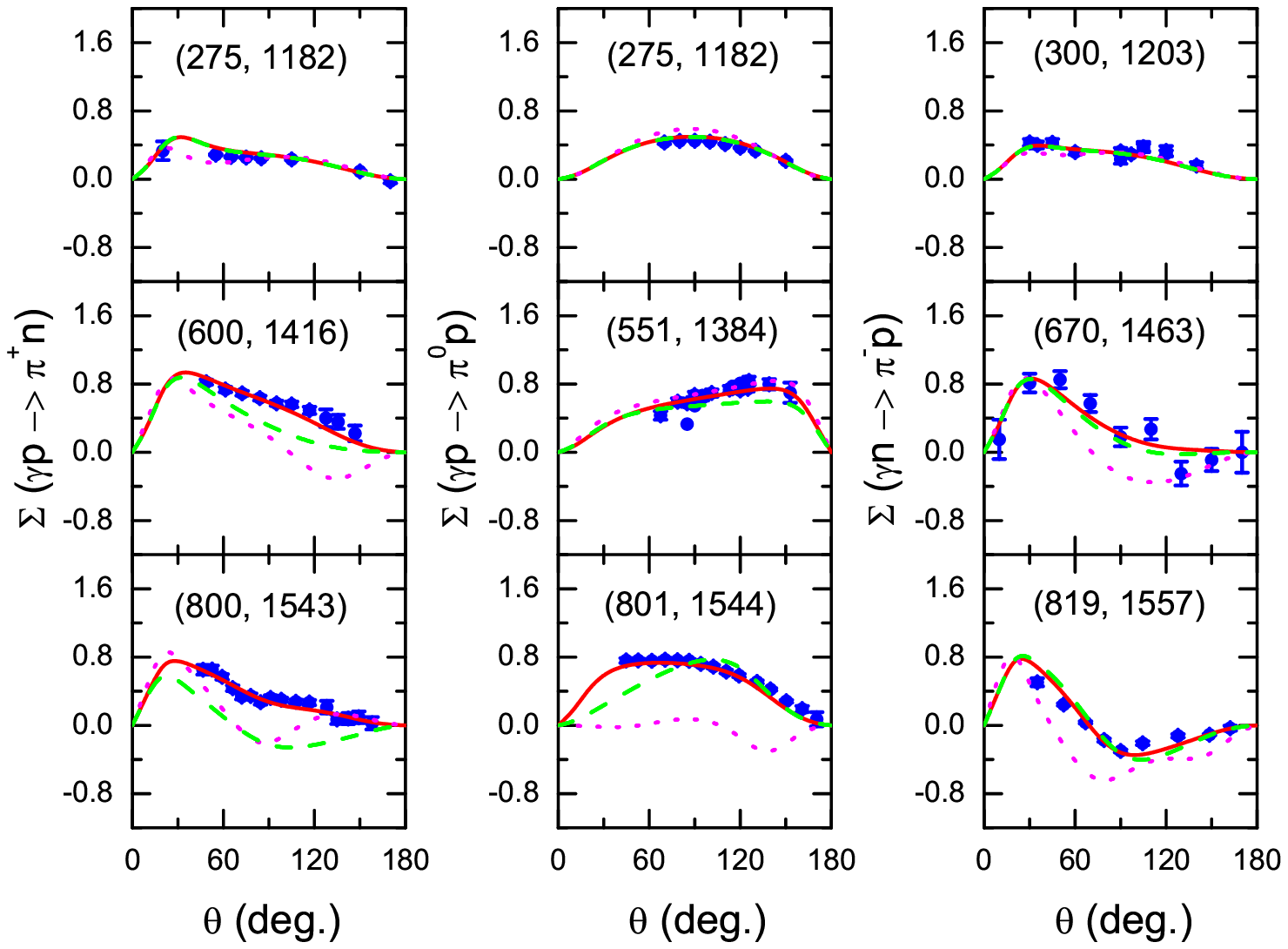}
\caption{\label{fig:sa_cc} (Color online) Photon spin asymmetry for $\gamma p
\to \pi^+ n$, $\gamma p\to \pi^0 p$ and $\gamma n \to \pi^- p$ as a function of
the scattering angle. The first and second number in each pair of parentheses
correspond to the photon laboratory momentum and the $\pi N$ c.m.\ energy,
respectively. The solid curves show the results from the full calculation. The
dotted curves and the dashed curves are obtained by respectively switching off
the $\pi\Delta$ loop integral and the $\eta N$ loop integral in
Eq.~(\ref{eq:MmuWithT}). Data are taken from
Refs.~\cite{Bartalini05,Elsner09,SAID}.}
\end{figure*}

\begin{figure*} [tb] \hglue 0.2cm
\includegraphics[width=0.68\textwidth,clip=]{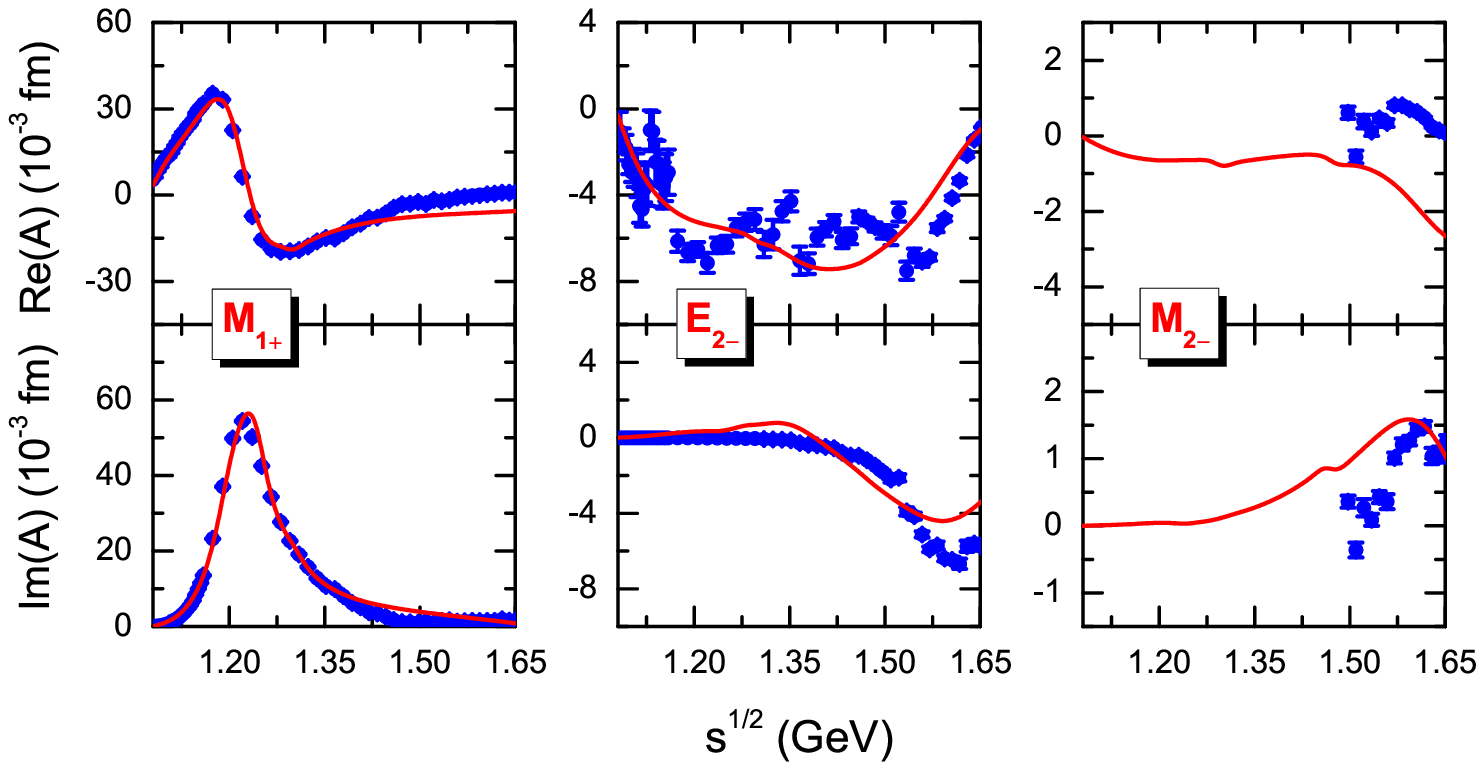}
\caption{\label{fig:multipole} (Color online) Real part and the imaginary part
of the multipole amplitudes $M_{1+}$, $E_{2-}$ and $M_{2-}$ for $\gamma N\to\pi
N$ with isospin $I=3/2$ as a function of the $\pi N$ c.m.\ energy. Scattered
symbols are amplitudes taken from the George Washington University's
partial-wave analysis \cite{Arndt02}.}
\end{figure*}

It was alluded to above in the context of the Bethe-Salpeter equation
(\ref{eq:TBSeq}) that the splitting of $T$ into pole and non-pole contributions
is no longer unique if the respective pieces are evaluated in a framework that
truncates the non-linearities. Owing to this arbitrariness, the pole and
non-pole parts of the J\"ulich model both exhibit spikes in the $P_{11}$
partial wave near the $\pi N$ threshold due to the presence of a nearby
unphysical pole in the non-pole transition matrix $X$. This pole in $X$ also affects the pole part of $T$  since the dressed vertices $F$ appearing in the splitting
(\ref{eq:TXsplit}) also depend on $X$ via Eq.~(\ref{eq:dressedF}), and one
finds that in the sum of pole and non-pole parts, these spikes cancel each
other precisely to yield a smooth full $T$-matrix \cite{Doring09L}. The spikes,
therefore, are unphysical artifacts of the particular way the pole and non-pole
contributions are calculated in the J\"ulich model. To circumvent problems of
this kind, we have chosen here, in Eq.~(\ref{eq:MmuWithT}), to work with the
full $T$-matrix to describe the hadronic final-state interaction. This,
however, does not eliminate completely all numerical difficulties because the
$s$-channel contribution in Eq.~(\ref{eq:MmuWithT}) still involves the dressed
$NN\pi$ vertex $F_s$ which, as explained, also exhibits the spike by itself. To
avoid this problem, we replace here the dressed vertex $F_s$ of the J\"ulich
model by the $NN\pi$ vertex obtained within the Feynman prescription for the
effective $NN\pi$ interaction Lagrangian ${\cal L}_{NN\pi}$  given in the
Appendix employing the physical $NN\pi$ coupling constant and physical nucleon
mass.

\section{Meson-baryon channels included in the present model} \label{sec:furtherdetail}

In the present work, as a first step towards a more complete calculation, only
the $\pi N$, $\eta N$ and $\pi \Delta$ channels are included as intermediate
states in the loop integral in Eq.~(\ref{eq:MmuWithT}). However, this
restriction only concerns the photoproduction sector. In the hadronic sector,
all of the five channels $\pi N$, $\eta N$, $\pi \Delta$, $\sigma N$, and $\rho
N$ are included via the channels that couple into all of the J\"ulich
$T$-matrices. The main reason we do not include the $\sigma N$ and $\rho N$
channels of the hadronic J\"ulich model in the present photoproduction
calculation is purely practical, namely, the corresponding transition couplings
of the photon to these channels would require additional free parameters
beyond what we can handle numerically at the moment. Moreover, for the $\rho N$
channel, we expect significant contributions only at energies higher than the
c.m.\ energy of $W=1.65$\,GeV considered in the present work.

The channels and intermediate hadrons that contribute explicitly in the present
calculation are shown in Fig.~\ref{fig:channels}. As depicted in
Fig.~\ref{fig:channels}(a), in addition to the nucleon that is treated as
described in Sec.~\ref{sec:III_Nucleon}, there are eight resonant contributions
that enter the $s$-channel term of Eq.~(\ref{eq:MmuWithT}), i.e.,
$S_{11}(1535)$, $S_{11}(1650)$, $S_{31}(1620)$, $P_{31}(1910)$, $P_{13}(1720)$,
$D_{13}(1520)$, $P_{33}(1232)$, and $D_{33}(1700)$. All of the corresponding
hadronic vertices $F_s$ and the dressed resonance propagators $S$ are taken
from the J\"ulich $\pi N$ model.  The corresponding electromagnetic
nucleon-resonance transition vertices $\tJ^\mu_s$ should be calculated from the
analog of Fig.~\ref{fig:NcurrentC}(b). However, since the transition currents
are all transverse by themselves, the analogs of the longitudinal loops in
Fig.~\ref{fig:NcurrentC}(b) vanish. Moreover, since we do not consider bare
Kroll-Ruderman-type four-point currents for the resonances, we are only left
with the bare currents that are given by the corresponding effective
Lagrangians (\ref{1hfNr}) and (\ref{3hfNr}) in the Appendix. This is one of the
gratifying features of the present formulation that greatly simplifies the
calculation of the $s$-channel resonant contributions.

The $\gamma N \to \pi N$ contributions of Fig.~\ref{fig:channels}(b) enter both
the Born-type current $B^\mu$ and the FSI loop $TG_0B^\mu_\tra$ of
Eq.~(\ref{eq:MmuWithT}). The contributions of Fig.~\ref{fig:channels}(c) and
Fig.~\ref{fig:channels}(d) for $\gamma N \to \eta N$ and $\gamma N \to \pi
\Delta$, on the other hand, only enter the FSI loop. For the $t$-channel
currents, we include the contributions from the $\pi$, $\rho$, $\omega$, and
$a_1$ exchanges in the $\gamma N \to \pi N$ channel, from $\rho$ and $\omega$
exchanges in the $\gamma N\to \eta N$ channel and from $\pi$ and $\rho$
exchanges in the $\gamma N\to \pi \Delta$ channel. These diagrams are
calculated using the corresponding Lagrangians listed in the Appendix. All the
hadronic coupling constants are taken from the J\"ulich $\pi N$ model
\cite{Janssen96,Schutz98,Krehl00,Gasparyan03}. The $MM'\gamma$ electromagnetic
coupling constants are determined from the radiative decay of the relevant
mesons in conjunction with SU(3) symmetry considerations. The numerical values
of the coupling constants are given the Appendix in connection with their
associated interaction Lagrangians. The only adjustable parameters here are the
cutoff parameters in the off-shell form factors introduced at the
electromagnetic vertices. These form factors are also given in the Appendix.

The only contribution to the $u$-channel amplitude $M^\mu_u$ considered in the
present work is that involving the nucleon (nucleonic current) and $\Delta$ in
the intermediate state. Contributions to the $u$-channel from the baryon
resonances are not included for consistency reasons since the J\"ulich hadronic
model (and most other existing dynamical models for that matter) does not
include the resonance contributions to the $u$-channel except the
$\Delta(1232)$ resonance. Apart from an enormous numerical demand in keeping
the self-consistency between the $s$- and $u$-channel amplitudes due to the
dressed vertices and propagators, the $u$-channel contribution from the baryon
resonances are not expected to introduce a significant energy dependence on the
resulting reaction amplitude.

We mention here that finally we are left with $38$ adjustable parameters in the
present work. Their values will be given and discussed in the next section.

\section{Results and discussion} \label{sec:results}

The photoproduction reactions of both the neutral and charged pions are
considered in the present work up to a c.m.\ energy of $W=1.65$ GeV. The free
parameters of the model as specified in the previous sections are determined by
fitting the available differential cross section and photon spin asymmetry
data. They are given in Tables~\ref{tab:para_cut}--\ref{tab:para_hat}. Here, we
note that in the present work, we have attached a momentum cutoff in the loop
integral in Eq.~(\ref{eq:MmuWithT}) of the form $\Lambda^2/(\Lambda^2 +
\vec{q}^{\,2})$, where $\vec{q}^{\,2}$ stands for the loop momentum and
$\Lambda$ is an adjustable cutoff parameter whose value is also given in Table~\ref{tab:para_cut}.

The resulting parameters of the effective baryon-resonance electromagnetic
transition vertices associated with the current $\tilde{J}^\mu_s$ as explained
in the previous section are summarized in Table~\ref{tab:para_res_1hf} for
isospin-1/2 resonances and in Table~\ref{tab:para_res_3hf} for isospin-3/2
resonances, respectively. We emphasize here that the given coupling constant
values should not be confused with the corresponding physical couplings, for
they are associated with the current $\tilde{J}^\mu_s$ [cf.
Eq.~(\ref{eq:MmuWithT})] which is only a part of the full current $J^\mu$ given
by Eq.~(\ref{eq:JmuDetail}). More appropriate (physical) resonance parameters,
i.e., masses and coupling strengths, should be associated with the poles of the
reaction amplitude, and their corresponding residues, in the complex energy
plane. The pole positions and hadronic residues have already been extracted
\cite{Doring09} in the J\"ulich hadronic reaction model. Efforts to extract the
resonance electromagnetic coupling strengths from the present model are in
progress and will be reported elsewhere.

The other adjustable parameters of the model, i.e., the cutoffs in the form
factors [Eqs.~(\ref{eq:ff}) and (\ref{eq:secondff})] and the parameters $C_1$,
$C_2$, and $\kappa_0$ in the nucleon electromagnetic current in
Eq.~(\ref{eq:newpar3}) are given in Table~\ref{tab:para_CCk}, while the
parameters $\hat{h}$ and $\beta$ in the contact current $M^\mu_c$ [cf.\
Eqs.~(\ref{eq:Fhat}) and (\ref{eq:Mcapprox})], are summarized in
Table~\ref{tab:para_hat}. All adjustable parameters were determined by global fits
that included all data sets as shown in Figs.~\ref{fig:dsig_pi+n}-\ref{fig:sa_pi0n}.

The calculated differential cross sections as a function of pion scattering
angle in the c.m.\ frame are shown in Figs.~\ref{fig:dsig_pi+n}--\ref{fig:dsig_pi0n} for $\gamma p\to \pi^+ n$, $\gamma p\to \pi^0 p$, $\gamma n\to \pi^- p$, and $\gamma n\to \pi^0 n$,
respectively, together with the corresponding data.  Similarly, the calculated
photon spin asymmetries are shown in Figs.~\ref{fig:sa_pi+n}--\ref{fig:sa_pi0n}. One sees that the overall agreement with the recent experimental data is very good. Some noticeable discrepancies
are seen in both the cross sections and beam asymmetries, the latter in $\gamma p \to \pi^0 p$ and $\gamma p \to \pi^+ n$ at higher energies. At this point, it is difficult to say whether these discrepancies are due to the lack of higher-spin resonances in the present model or due to the coupled-channels effects other than those from $\eta N$ and $\pi \Delta$ in the photoproduction
kernel. In this connection, we mention that the $K\Sigma$ channel with isospin $3/2$ has been incorporated into the J\"ulich model quite recently \cite{RDSHM10} and the inclusion of the isospin-1/2 $K\Sigma$ and $K\Lambda$ channels are currently in progress. The inclusion of the $\rho N$, $\sigma N$ and strangeness channels as well as higher spin resonances such as the $D_{15}$ and $F_{15}$ resonances into our model requires extra free parameters and will be done in future work, which may give us a chance to improve the quality of the present description of the experimental data, especially, in the $W \sim 1.6$ GeV region.

Another multi-channel dynamical model available in the literature which
analyzes the pion photoproduction reactions is the EBAC model \cite{Diaz08}.
This model, based on a unitary transformation method, includes two more
resonances than the present model, namely the $D_{15}(1675)$ and $F_{15}(1680)$
resonances; it also includes two more channels in the intermediate state for
the photoproduction process, namely the $\rho N$ and $\sigma N$ channels. On
the other hand, it analyzes neither the $\gamma n \to \pi^- p$ nor the $\gamma
n \to \pi^0 n$ reactions. Although they have considered older data than those
in the present work, thus making a close comparison difficult, their results
are of comparable fit quality to the present model results overall for both the
$\gamma p \to \pi^0 p$ and $\gamma p \to \pi^+ n$ reactions. The present
results are slightly better at higher energies, especially for photon spin
asymmetries for the $\gamma p \to \pi^0 p$ reaction. Anyway, to achieve the
level of the fit quality of the EBAC model results, no spin-5/2 resonances are
required in the present model calculation of the pion photoproduction reaction.
Further studies are needed to understand the role of those resonances.

At low energies, close to threshold, the pion photo- and electroproduction
reactions are nowadays completely understood thanks to ChPT
\cite{Bernard92,BKLM94}.  Any meson-exchange dynamical model should, in
principle, have built in the constraints of ChPT. It is, however, not a simple
task to account for all the constraints dictated by ChPT and, in general, only
a few basic constraints are taken into account in practice. Indeed, building in
the chiral constraints into meson-exchange models is one of the major
improvements needed for these models. The J\"ulich model uses the
phenomenological chiral Lagrangian of Wess and Zumino \cite{Wess67},
supplemented by additional Lagrangians for the coupling of $\Delta$, $\omega$,
$\eta$, $a_0$, and $\sigma$ \cite{Krehl00,Gasparyan03}, thus honoring some of
the chiral constraints. However, it also contains phenomenological form factors
which spoil these constraints in general.

Nevertheless, it is interesting to see how the present meson-exchange dynamical
model performs at low energies. In Fig.~\ref{fig:dsdo_th}, we show our
differential cross section results (solid lines) at low energies close to
threshold, together with the experimental data. These results, however, were
obtained by a refit of the parameters $\hat{h}$ and $\beta$ of the generalized
contact current (\ref{eq:Mcapprox}), leaving all other parameters unchanged,
since it was not possible to obtain reasonable fits using the same constants as
for the global fit, in particular, for the $\pi^0 p$ channel. The refitted values of ($\hat{h}$, $\beta$) are ($12.44$, $0.04$) for the $\pi^+ n$ channel,  ($4.13$, $-1.72$) for the $\pi^0 p$
channel, and ($6.58$, $-0.83$) for the $\pi^- p$ channel, respectively.
Comparing with those values listed in Table~\ref{tab:para_hat}, one sees that
these values refitted to the data close to threshold are quite different. This
should not be surprising in view of the fact that, in principle, these
parameters are energy-dependent functions that are being treated here as
constants for simplicity. Near threshold, this restriction becomes especially
noticeable. One sees that the overall agreement is reasonable, except for the
normalization of the differential cross sections for $\gamma p \to \pi^0 p$ at
$W=1084$ MeV and for $\gamma p \to \pi^+ n$ at $W=1080$ MeV.

One should also keep in mind that the present calculation is performed in the
isospin basis, with averaged nucleon and pion masses of $938.92$ and $138.04$
MeV, respectively, which corresponds to a common threshold energy of about
$1077.0$ MeV. This value is about $2$ MeV below the correct threshold value of
$1079.1$ MeV for the $\gamma p \to \pi^+ n$ reaction and might explain, at
least in part, why the present calculation over-predicts the cross section at
$W=1080$ MeV which is just about $1$ MeV above the correct threshold value. In
fact, the dashed curve in the corresponding panel represents the result
obtained with the photo-transition amplitudes $F_s S\tilde{J}^\mu_s$ and
$B^\mu$ [cf. Eqs.~(\ref{eq:MmuWithT}) and (\ref{eq:BornCurrent})] calculated
with the averaged mass of the nucleon and pion set to the neutron and $\pi^+$
mass value, respectively, while the hadronic scattering amplitudes are still
calculated in the isospin basis. One sees that the agreement with the data is
now improved to some extent at $W=1080$ MeV, but not at a higher energy of
$W=1112$ MeV. Similarly, the common threshold energy value is about $3.8$ MeV
above the correct threshold value of $1073.2$ MeV for the reaction $\gamma p
\to \pi^0 p$. Therefore, in this case we expect the calculated cross section to
under-predict the data. Indeed, as can be seen in Fig.~\ref{fig:dsdo_th} (solid
curve), this is the case even for the data at $W=1084$ MeV which is about
$10.8$ MeV above the correct threshold value. The dashed curve here corresponds
to the result analogous to that for the $\pi^+ n$ final state but with the
averaged mass of the nucleon and pion set to the proton and $\pi^0$ mass,
respectively. One sees a non-negligible improvement in the agreement with the
data. In the $\gamma n \to \pi^- p$ reaction, the common threshold value is
about $0.8$ MeV below the correct value of $1077.8$ MeV and is much smaller
than the corresponding differences in the other two reactions just discussed.
Therefore, in this case we do not expect that such a small energy difference
will affect much the predicted results, especially, if we are at energies not
so near the threshold as at $W=1085$ MeV shown in Fig.~\ref{fig:dsdo_th}. The
dashed curve here is the analog of those for $\pi^+ p$ and $\pi^0 p$ for the
case of the $\pi^- p$ final state, and the effect is negligible as expected. We
thus conclude that, apart from the isospin symmetry violation effects arising
from the mass differences of pions and nucleons, the present model works
remarkably well even at low energies close to threshold. A closer comparison
with the low-energy data, however, requires a calculation in the full particle
basis. This is especially true for the reaction $\gamma p \to \pi^0 p$, where
it is well known that the near-threshold cross section results from
cancelations of competing mechanisms that yield relatively large contributions
individually and, therefore, making the prediction of its correct threshold
behavior a non-trivial issue \cite{Bernard92,Bernard:1994gm,DN10a}.

The influence of the loop integral in Eq.~(\ref{eq:MmuWithT}) and the
generalized contact current $M^\mu_c$ other than the Kroll-Ruderman term in
Eq.~(\ref{eq:BornCurrent}) on the differential cross sections and photon spin
asymmetries are illustrated in Figs.~\ref{fig:dsdo} and \ref{fig:sa},
respectively. There, the solid curves correspond to the results of the full
calculation. The dotted curves are obtained by switching off the loop integral
[the term proportional to the hadronic amplitude $T$ in
Eq.~(\ref{eq:MmuWithT})]. As one can see, the contribution from the loop
integral (which provides the effect of the hadronic final-state interaction) to
the cross sections and beam asymmetries is very important showing that it
cannot be ignored in these reactions. Note that, in this work, we have the full
$T$ in the loop integral of Eq.~(\ref{eq:MmuWithT}) instead of the non-pole
amplitude $X$ employed in the earlier feasibility
study of Ref.~\cite{Haberzettl06}. The dashed curves are obtained by switching
off all the terms except the usual Kroll-Ruderman contact term in the
generalized contact current $M^\mu_c$ of Eq.~(\ref{eq:Mcapprox}). As discussed,
these contact terms, unique to the present approach, are required to maintain
gauge invariance of the reaction amplitude. Leaving them out, significant
effects are seen on the cross sections in both the $\gamma p \to \pi^+ n$ and
$\gamma p \to \pi^0 p$ processes;  the effect in the $\gamma n \to \pi^- p$
process is negligible. These gauge-invariance-preserving terms also influence
the beam asymmetries at higher energies in a non-trivial way. These
interaction-current contributions are seen to be just as important here as they
were found to be for $NN$ bremsstrahlung in Ref.~\cite{NH09}, as discussed in
the Introduction (see also the Summary, Sec.~\ref{sec:summary}).
The results for both reactions clearly demonstrate that the gauge
invariance required for maintaining the internal consistency of the full
amplitude in a microscopic model \textit{is not} purely a theoretical issue
but is necessary for the description of the consistent reaction dynamics
(see Ref.~\cite{HN10} for a more detailed discussion on this point).

Figure~\ref{fig:sig} shows the total cross section as a function of the $\pi N$
c.m.\ energy for the $\gamma p \to \pi^+ n$, $\gamma p\to \pi^0 p$ and $\gamma
n \to \pi^- p$ reactions. As one can see, the agreement of the full results
(solid curves) with the data --- the latter taken from Ref.~\cite{SAID} were
not included in the global fit --- is remarkably good over the entire energy
range considered. This figure also shows the influence of the FSI contributions
(dotted curves). Again, they are seen to be very important. The influence of
the contact terms apart from the Kroll-Ruderman term in the generalized contact
current $M^\mu_c$ [Eq.~(\ref{eq:Mcapprox})] is also shown as dashed curves.
Here, a significant effect is seen in the $\gamma p \to \pi^+ n$ reaction in
the energy range $W \geq 1.3$ GeV and in the $\gamma p \to \pi^0 p$ reaction
around the energy $W \sim 1.2$ GeV. Note for the $\gamma p \to \pi^0 p$
reaction, the effect on the total cross section is largely suppressed as
compared to that on the differential cross sections (see Fig.~\ref{fig:dsdo})
due to the phase space factor $\sin(\theta)$.

The effects of the photon coupling to the $\pi\Delta$ and $\eta N$ channels on
the cross sections and beam asymmetries are illustrated in
Figs.~\ref{fig:dsdo_cc} and \ref{fig:sa_cc}, respectively. There, the solid
curves correspond to the full calculation; the dotted curves and dashed curves
are obtained by switching off, respectively, the $\pi\Delta$ and $\eta N$
channels in the loop integral in Eq.~(\ref{eq:MmuWithT}). Note that the
threshold energy for opening of the $\pi\Delta$ channel is at  $W\cong 1.37$
GeV, while for the $\eta N$ channel is at $W\cong 1.49$ GeV. One sees that the
effect of the $\eta N$ channel is practically negligible for cross sections,
while it shows some influence on the beam asymmetries at higher energies. The
effect of the $\pi\Delta$ channel is significant for reproducing both the cross
section and beam asymmetry data at higher energies. The reason for this is that
there is an efficient overlap under the loop integral between the generalized
contact current $M^\mu_c$ and the $\pi\Delta \to \pi N$ transition amplitude in
the $D_{31} \to S_{31}$ partial-wave state which peaks around $1.6$ GeV
due to the strong coupling of the $S_{31}(1620)$ resonance to the $\pi\Delta$
channel. There is also a significant contribution from the $D_{11} \to S_{11}$
partial wave due to the $S_{11}(1535)$ resonance. Here, it should be noted
that, since so far the hadronic model parameters in the $\pi\Delta$ channel
have not been constrained by the data, the strong coupling of these resonances to the
$\pi\Delta$ channel might be an artifact of the model in the hadronic sector.
While the present photoproduction reaction may help constrain some of the $\pi \Delta$-channel parameters, more conclusive results can be expected from investigations
of two-pion productions $\pi N \to \pi\pi N$ and  $\gamma N \to \pi\pi N$.

Motivated by the good agreement of our results with the total and differential
cross sections as well as the beam asymmetry data, we have extracted the
multipole amplitudes for pion photoproduction and compared them with those from
the George Washington University's partial-wave analysis \cite{Arndt02}. In
Fig.~\ref{fig:multipole}, the results for the multipole amplitudes $M_{1+}$,
$E_{2-}$, and $M_{2-}$ from the present calculation (solid curves) are shown
together with the results from the SAID analysis \cite{Arndt02}. The latter
includes not only the cross sections and beam asymmetries but also the target
asymmetries and recoil nucleon polarizations as well as some double
polarization data into its analysis. We see that the agreement between the two
results for the dominant $M_{1+}$ amplitude is quite good. For the smaller
$E_{2-}$ amplitude, the agreement is also reasonable, but for the small
$M_{2-}$ amplitude there is a considerable disagreement. This illustrates the
kind of uncertainties one should expect from the present-type calculations,
even though we reproduce the cross sections and beam asymmetries quite nicely.
It is clear that in order to extract more reliable multipoles (apart from the
dominant ones) from the present model, one needs to include more independent
observables to further constrain the model. Actually, the SAID results are also
subject to some assumptions in their analysis since at present there exists no
complete set of data. Indeed, in order to uniquely determine the amplitude in
the present reaction, one requires at least eight independent observables
\cite{Tabakin} (See, also a recent discussion \cite{Tiator11} on this issue).
In a recent analysis \cite{Workman10}, Workman has also investigated the
sensitivity of the extracted multipole amplitudes to the accuracy of the data
used in their extraction.

\section{Uncertainties} \label{sec:uncertainties}

Regarding the assessment of theoretical uncertainties of our results, this is very
difficult to do in a quantitatively reliable manner within the present
phenomenological effective Lagrangian approach because of the absence of a
precise ordering scheme for refining the approximations. A procedure was
outlined in Ref.~\cite{RDSHM10} for pion-nucleon scattering that allows one to
quantify how the error margins of the data carry over into uncertainties of the
extracted parameters of the model approach. A similar approach could be used as
well to assess the statistical errors of the photoproduction reaction. However,
at present, no quantitative error analysis is available for the J\"ulich model
that we employ here for the hadronic final-state interaction and so we must
postpone such an investigation to future work.

In addition to the statistical errors, there are systematic uncertainties
inherent in all phenomenological effective Lagrangian approaches that stem from
the implementation (or violation) of Lorentz covariance, unitarity,
analyticity, and (for photoprocesses) gauge invariance, from the truncation of
reaction channels and from how many intermediate resonances are taken into
account. For the present approach, we expect the last two sources of
uncertainties to be most relevant. Plans are underway to address these issues
in future applications by including more channels ($\omega N$, $K\Lambda$,
$K\Sigma$) and higher-spin resonances [such as $D_{15}(1675)$ and
$F_{15}(1680)$]. However, it may well be that a better understanding of the
systematic errors of phenomenological effective Lagrangian approaches can only
be obtained by comparing the results of different formalisms. As an example, we
point here to the fact that while the present results, by and large, are of a
quality similar to that of the EBAC model \cite{Diaz08}, we do not need the spin-5/2 resonances employed by that model. This is an important finding in itself since
it places the actual role of these resonances in check. However, it may also be
an indication that these resonances are needed in the EBAC model to make up for
some basic deficiencies of that model, e.g., lack of Lorentz covariance and gauge invariance. In any case, this points to the necessity for further investigating such higher-spin resonances to get a better understanding of the corresponding systematic uncertainties, something we shall do in the future.

\section{Summary} \label{sec:summary}

We have presented results for the neutral and charged pion
photoproduction reactions within a coupled-channels dynamical model in
conjunction with the J\"ulich $\pi N$ hadron-exchange model. The
photoproduction amplitude in the present approach satisfies the important
properties of analyticity, unitarity and gauge invariance, the latter as
dictated by the generalized Ward-Takahashi identity. The overall
agreement with the cross section and beam asymmetry data of these reactions is
very good in the entire energy range considered. Even at very
low energies close to threshold, we have shown that, apart from the
isospin-symmetry violation arising from the mass differences of pions
and nucleons, the present model works quite well, especially, in view of the
delicate cancelations among various competing mechanisms in the $\gamma p \to
\pi^0 p$ reaction near threshold. A closer comparison with the data close to
threshold, however, requires a calculation in the particle basis.

The present model includes only the spin-1/2 and -3/2 resonances, showing that,
within this model, there is no obvious indication for the need of higher spin
resonances --- in particular, the $D_{15}(1675)$ and $F_{15}(1680)$ resonances
--- in the energy range of up to 1.65 GeV to describe the pion photoproduction
cross section and beam asymmetry data. In this connection, it is very
interesting to extend the present calculation to other (spin) observables in
pion photoproduction.

The appearance of the terms in addition to the usual Kroll-Ruderman contact
term in the generalized contact current $M^\mu_c$ [Eq.~(\ref{eq:Mcapprox})] is a
unique feature of the present model. These terms account for the complicated
parts of the interaction current that cannot be taken into account explicitly
at present. Our results show that these terms have significant effects on the
calculated observables in the present reactions. This means it is very import
to take into account properly the gauge-invariance-preserving interaction
current for the pion photoproduction processes. The importance of this current
corroborates similar findings reported for the $NN$ bremsstrahlung reaction
quite recently \cite{NH09}, where it was found to be crucial in
reproducing the KVI data \cite{KVI02}, something which had eluded theoretical
attempts for a very long time. The important point here is that this was
brought about simply by adding the gauge-invariance-preserving current
$M^\mu_c$, as it is determined here for pion photoproduction, as a novel
four-point contact-current mechanism  into the description of $NN$
bremsstrahlung, \textit{without changing any of the other mechanisms for that
reaction}. This reciprocal consistency between the current mechanisms employed
in the two processes clearly demonstrates that maintaining gauge
invariance of the reaction amplitudes, as dictated by the
respective generalized Ward-Takahashi identities of all contributing current mechanisms, is not a purely theoretical issue, but an indispensable requirement for a
consistent, correct description of the reaction dynamics, with direct consequences for our ability to reproduce the experimental data \cite{NH09,HN10}.

Apart from the dominant multipole amplitudes, the smaller multipole amplitudes extracted
from the present model calculation are shown to be subject to considerable
uncertainties, even though the model reproduces quite nicely the recent cross
section and beam asymmetry data. It is clear that other independent spin
observables need to be included in the analysis to further
constrain the model. Note that a unique determination of the multipole
amplitudes in pion photoproduction requires, in principle, at least eight
independent observables \cite{Tabakin} which are not available at
present.

As we have emphasized in the previous section, the nucleon-resonance
electromagnetic transition couplings displayed in Tables~II-III
are not the physical coupling values and, as such, they are associated only
with the present calculation. The appropriate physical
electromagnetic couplings should be extracted from the residues associated with
the poles of the photoproduction amplitude in the complex-energy plane. The
work in this direction is underway and the results will be reported elsewhere.

Finally, we have considered the $\pi N$ c.m.\ energies up to 1.65 GeV. This
upper limit is set by the limitation of the J\"ulich hadronic model we employed
here for the hadronic interactions. To analyze data at higher energies, one
needs to include higher-spin baryon resonances, such as $D_{15}(1675)$ and
$F_{15}(1680)$. Also, one needs to perhaps also include other meson-baryon
channels, such as the $K\Lambda$, $K\Sigma$, and $\omega N$ channels. The
$K\Sigma$ channel with isospin $3/2$ has just been incorporated \cite{RDSHM10}
into the J\"ulich hadronic model and the inclusion of the strangeness channels
with isospin $1/2$ is in progress. The $K\Lambda$ and $K\Sigma$ channels are
expected to play a particularly important role in $\eta$ photoproduction
\cite{DN10}.

In summary, the present work provides a comprehensive treatment of pion
photoproduction within a covariant coupled-channels framework based on
phenomenological effective Lagrangians. The details of the approach have been
constructed \cite{HHN2011} with two main goals in mind, namely to preserve the
gauge invariance of the current as an off-shell condition and to allow for the
consistent incorporation of the hadronic final-state interaction. Overall, our
results show very good agreement with the data and they, moreover, show that
both properties are indispensable if one wants to provide a quantitatively
reliable description of the reaction dynamics of pion photoproduction across
the entire resonance region. The extraction of the electromagnetic
transition coupling constants for resonances from the residues associated with
the poles of the reaction amplitude is currently in progress. It is
also straightforward to extend the present approach to the production
of other mesons, to strangeness production, and also to the electroproduction
of mesons. Tackling all of these reactions is planned for the near future.

\begin{acknowledgments}
The authors are indebted to Ashot Gasparyan for his help in providing the
necessary ingredients from the J\"ulich hadronic model in the early stage of
this work. We also thank Shan-Ho Tsai, Bruno Juli\'a-D\'iaz, Mark Paris, and
Andreas Nogga for their help with parallel programming aspects. This work is
supported by the FFE grant No. 41788390 (COSY-058). The work of M.D. has been
supported by the DFG (Deutsche Forschungsgemeinschaft, GZ: DO 1302/1-2)
and the EU Integrated Infrastructure Initiative HadronPhysics2
(contract No. 227431). The authors acknowledge the Georgia Advanced Computing
Resource Center at the University of Georgia and the J\"ulich Supercomputing
Center at Forschungszentrum J\"ulich (Project ID jikp07) for providing
computing resources that have contributed to the research results reported
within this paper.
\end{acknowledgments}

\appendix*

\section{Lagrangians and form factors}

We list here the Lagrangians and form factors used in the present work.

The hadronic interaction Lagrangians are:
\begin{align}
{\cal L}_{NN\pi} &=  -\, g_{NN\pi}  {\bar N}  \left[ \gamma_5
\left( i\lambda + \frac{1 - \lambda}{2M_N}\,
\dslash{\partial} \right) {\bm\pi} \cdot {\bm\tau} \right] N , \label{NNpi} \\[1pt]
{\cal L}_{NN\eta} &=  -\, g_{NN\eta}  {\bar N}  \left[ \gamma_5
\left( i\lambda + \frac{1 - \lambda}{2M_N}\,
\dslash{\partial} \right) \eta \right] N , \label{NNeta} \\[1pt]
{\cal L}_{NN\rho} &=  -\, g_{NN\rho}  {\bar N} \left[ \left(
\gamma^\mu - \frac{\kappa_\rho}{2M_N}\sigma^{\mu\nu}\partial_\nu
\right) {\bm \rho}_\mu \cdot {\bm\tau} \right] N, \label{NNrho} \\[1pt]
{\cal L}_{NN\omega} &=  -\, g_{NN\omega}  {\bar N} \left[ \left(
\gamma^\mu - \frac{\kappa_\omega}{2M_N}\sigma^{\mu\nu}\partial_\nu
\right)
\omega_\mu \right] N, \label{NNomega} \\[3pt]
{\cal L}_{NNa_1} &= \, g_{NNa_1} {\bar N} \gamma^\mu\gamma_5 {\bm
\tau}\cdot\bv{a}_1^\mu N, \label{NNa1} \\[3pt]
{\cal L}_{\Delta N\pi} &= \, \frac{g_{\Delta N\pi}}{m_\pi} \bar{\Delta}^\mu {\bm T} \partial_\mu {\bm \pi} N + {\rm H.c.}, \label{DNpi} \\[3pt]
{\cal L}_{\Delta N\rho} &= -\,i\,\frac{g_{\Delta N\rho}}{2m_N} {\bar{\Delta}^\mu} \gamma^\nu \gamma_5 {\bm T} {\bm \rho}_{\mu\nu} N + {\rm H.c.}, \label{DNrho}
\end{align}
where $\lambda$ in Eqs.~(\ref{NNpi}) and (\ref{NNeta}) is the mixing parameter
of the pseudoscalar ($\lambda=1$) and pseudovector ($\lambda=0$) type
couplings. In this work, $\lambda$ is taken to be zero which means we adopt the
pure pseudovector type coupling. The coupling constant values in the above
Lagrangians are given in Table~\ref{tab:para_0}. All those values have also been
used for the hadronic part of the amplitude \cite{Krehl00,Gasparyan03} employed
in the present work.

\begin{table}[t!]
\caption{\label{tab:para_0} Coupling constant values of the present model fixed from independent sources. The references are shown in the third column.
See text for details. $m_{a_1}=1260$ MeV.}
\renewcommand{\arraystretch}{1.3}
\begin{tabular*}{\columnwidth}{@{\extracolsep\fill}crr}
\hline\hline
 $g_{NN\pi}  $           &  $13.46$       & \cite{Schutz98}   \\
 $g_{NN\eta}$            &  $4.76$        & \cite{Schutz98}   \\
 $g_{NN\rho}$            &  $3.25$        & \cite{Janssen96}  \\
 $\kappa_\rho$           &  $6.10$        & \cite{Janssen96}  \\
 $g_{NN\omega}$          &  $11.76$       & \cite{Janssen96}  \\
 $\kappa_\omega$         &  $0.00$        & \cite{Janssen96}  \\
 $g_{NNa_1}$             &  $\dfrac{m_{a_1}}{m_\pi}g_{NN\pi}$  & \cite{Wess67}    \\
 $g_{\Delta N\pi}$       &  $ 2.13$       & \cite{Janssen94,Janssen942,Machleidt87} \\
 $g_{\Delta N\rho}$      &  $-39.10$      & \cite{Janssen94,Janssen942,Machleidt87} \\
 $g_{\gamma\pi\rho}$     & $0.11$         & \cite{NDHHS99,PDG,Garcilazo93}         \\
 $g_{\gamma\pi\omega}$   & $0.32$         & \cite{NDHHS99,PDG,Garcilazo93}         \\
 $g_{\gamma\eta\rho}$    & $0.89$         & \cite{NDHHS99,PDG,NOH08}               \\
 $g_{\gamma\eta\omega}$  & $0.25$         & \cite{NDHHS99,PDG,NOH08}               \\
 \hline\hline
\end{tabular*}
\end{table}

The electromagnetic interaction Lagrangians for the nucleon and mesons read
\begin{align}
{\cal L}_{NN\gamma} =& \, -e \bar N \left[ \left(\hat{e} \gamma^\mu
- \frac{\hat{\kappa}}{2M_N}\sigma^{\mu\nu}\partial_\nu \right) A_\mu
\right] N, \label{NNr} \\[2pt]
{\cal L}_{\gamma\pi\rho} =&\; e\frac{g_{\gamma\pi\rho}}{m_\pi}
\varepsilon_{\alpha\mu\lambda\nu} \left(\partial^\alpha
A^\mu\right) \left(\partial^\lambda {\bm\pi}\right) \cdot {\bm \rho}^\nu, \label{L_gpirho} \\[2pt]
{\cal L}_{\gamma\pi\omega} =&\; e\frac{g_{\gamma\pi\omega}}{m_\pi}
\varepsilon_{\alpha\mu\lambda\nu} \left(\partial^\alpha
A^\mu\right) \left(\partial^\lambda \pi_3\right) \omega^\nu, \label{L_gpiome} \\[2pt]
{\cal L}_{\gamma\eta\rho} =&\; e\frac{g_{\gamma\eta\rho}}{m_\eta}
\varepsilon_{\alpha\mu\lambda\nu} \left(\partial^\alpha
A^\mu\right) \left(\partial^\lambda \eta \right) \rho_3^\nu, \label{L_getarho} \\[2pt]
{\cal L}_{\gamma\eta\omega} =&\; e\frac{g_{\gamma\eta\omega}}{m_\eta}
\varepsilon_{\alpha\mu\lambda\nu} \left(\partial^\alpha
A^\mu\right) \left(\partial^\lambda \eta\right) \omega^\nu, \label{L_getaome} \\[2pt]
\mathcal{L}_{\gamma\pi\pi} =& \; e \left[\left(\partial_\mu {\bm
\pi}\right)\times {\bm \pi}\right]_3 A^\mu, \\[2pt]
\mathcal{L}_{\gamma\pi a_1} =& \; e \,\frac{1}{m_{a_1}}
F_{\mu\nu}\left[2\left(\partial^\mu{\bm \pi}\right)\times {\bm
a}_1^\nu \,-\, 2\left(\partial^\nu{\bm \pi}\right) \times {\bm
a}_1^\mu \right. \nonumber \\
& + \left. {\bm \pi}\times {\bm a}_1^{\mu\nu}\right]_3,
\end{align}
where $e$ stands for the elementary charge unit, and $\hat e \equiv (1 +
\tau_3)/2$ and $\hat\kappa \equiv \kappa_p(1 + \tau_3)/2 + \kappa_n(1 -
\tau_3)/2$, with the anomalous magnetic moments $\kappa_p=1.793$ of the proton
and $\kappa_n=-1.913$ of the neutron; $F_{\mu\nu}\equiv \partial_\mu A_\nu -
\partial_\nu A_\mu$ with $A_\mu$ denoting the electromagnetic field and ${\bm
a}_1^{\mu\nu}\equiv \partial^\mu {\bm a}_1^\nu -\partial^\nu {\bm a}_1^\mu$;
$\varepsilon_{\alpha\mu\lambda\nu}$ is the totally antisymmetric Levi-Civita
tensor with $\varepsilon^{0123}=+1$. The meson-meson electromagnetic transition
coupling constants in the above Lagrangians are given in Table~\ref{tab:para_0}.
Following Ref.~\cite{NDHHS99}, they are fixed from the
decay \cite{PDG} of the $\rho$ and $\omega$ meson into the $\gamma\pi^0$ and
$\gamma\eta$ channels, respectively. The signs of the coupling constants
$g_{\gamma\pi\rho}$ and $g_{\gamma\pi\omega}$ are consistent with those
determined from the study of pion photoproduction in the 1-GeV region
\cite{Garcilazo93}. The signs of $g_{\gamma\eta\rho}$ and
$g_{\gamma\eta\omega}$ are inferred from the flavor SU(3) symmetry
considerations as used in Ref.~\cite{NOH08}.

The resonance-nucleon photo-transition Lagrangians are
\begin{align}
\mathcal{L}^{1\!/2 \pm}_{RN\gamma} = & \;  e \frac{g_{RN\gamma}^{(1)}}{2M_N} \bar R \Gamma^{(\mp)}
\sigma_{\mu\nu}\left(\partial^\nu A^\mu\right) N + {\rm H.c.}, \label{1hfNr} \\[1pt]
\mathcal{L}^{3\!/2\pm}_{RN\gamma} = & \, -ie\frac{g^{(1)}_{RN\gamma}}{2M_N} {\bar{R}^\mu} \gamma_\nu
\Gamma^{(\pm)} F^{\mu\nu} N \nonumber \\
&\, + e\frac{g^{(2)}_{RN\gamma}}{4M^2_N} \bar{R}^\mu \Gamma^{(\pm)} F^{\mu\nu}
\partial_\nu N + {\rm H.c.}, \label{3hfNr}
\end{align}
where $\Gamma^{(+)}\equiv\gamma_5$ and $\Gamma^{(-)}\equiv 1$; the superscript
of $\mathcal{L}_{RN\gamma}$ denotes the spin and parity of the resonance $R$.

For the $s$-channel $NN\pi$ vertex and the $u$-channel $NN\pi$ and $\Delta
N\pi$ vertices, the following covariant form factor is employed in our model:
\begin{equation} \label{eq:ff}
F_B(p^2) = \left(\frac{\Lambda_B^4}{\Lambda_B^4+\left(m_B^2-p^2\right)^2}\right)^n,
\end{equation}
where $p$ and $m_B$ denote the four-momentum and mass of the off-shell baryon,
respectively. The exponent $n$ is taken to be $2$ for $\Delta N\pi$ vertex and $1$ for $NN\pi$ vertex. The parameter  $\Lambda_B$ is determined by fitting to the data and it is listed in Table~\ref{tab:para_cut}.

For the hadronic vertices in the $t$-channel diagrams, the following covariant
form factor is included
\begin{equation} \label{eq:secondff}
F_\alpha(q^2) =
\left(\frac{\Lambda_\alpha^2-m_\alpha^2}{\Lambda_\alpha^2-q^2}\right)^{n_\alpha},
\end{equation}
where $\alpha$ stands for the off-shell meson ($\alpha=\pi$, $\rho$, $\omega$,
$a_1$); $q$ and $m_\alpha$ denote the four-momentum and mass of the off-shell
meson, respectively. The exponent $n_\alpha$ is taken to be $1$ for
$\alpha=\pi$ \cite{Janssen94,Janssen942} and $2$ for other mesons
\cite{Machleidt87}. We use the same cutoff $\Lambda_v$ for $\rho$, $\omega$ and
$a_1$ mesons in order to reduce the number of model parameters. The values of
the cutoff parameters $\Lambda_\pi$ and $\Lambda_v$ are determined by fitting
to the data; they are listed in Table~\ref{tab:para_cut}.

Note that the gauge invariance feature of our photoproduction amplitude is
independent of the specific form of the form factors.


\begin{thebibliography}{99}
\bibitem{Klempt:2009pi}E. Klempt and J.M. Richard, Rev.\ Mod. Phys.\ \textbf{82}, 1095 (2010).
\bibitem{Cutkosky:1979fy}R.E. Cutkosky, C.P. Forsyth, R.E. Hendrick, and R.L. Kelly, Phys.\ Rev.\ D \textbf{20}, 2839 (1979).
\bibitem{Cutkosky80}R.E. Cutkosky, C.P. Forsyth, J.B. Babcock, R.L. Kelly, and R.E. Hendrick, in \textit{Baryons 1980, Proceedings of the IVth International Conference on Baryon Resonances}, ed.\ N. Isgur (University of Toronto, 1980), p.\ 19.
\bibitem{Hoehler83}G. H\"ohler, Pion-Nucleon Scattering, Landolt-B\"ornstein Vol.\
    \textbf{I/9b2}, ed.\ H. Schopper (Springer, Berlin, 1983).
\bibitem{Hohler93}G. H\"ohler, $\pi N$ Newsletter \textbf{9}, 1 (1993).
\bibitem{Arndt95}R.A. Arndt, I.I. Strakovsky, R.L. Workman, and M. Pavan, Phys.\ Rev.\ C \textbf{52}, 2120 (1995).
\bibitem{Arndt02}R.A. Arndt, W.J. Briscoe, I.I. Strakovsky, and R.L. Workman, Phys.\ Rev.\ C \textbf{66}, 055213 (2002).
\bibitem{Arndt06}R.A. Arndt, W.J. Briscoe, I.I. Strakovsky, and R.L. Workman, Phys.\ Rev.\ C \textbf{74}, 045205 (2006).
\bibitem{Bernard:1991rt}V. Bernard, N. Kaiser, J. Gasser, and U.G. Mei{\ss}ner, Phys.\ Lett.\  B \textbf{268}, 291 (1991).
\bibitem{Bernard92}V. Bernard, N. Kaiser, and U.-G. Mei{\ss}ner, Nucl.\ Phys.\ B \textbf{383}, 442 (1992).
\bibitem{Bernard:1994gm}V. Bernard, N. Kaiser, and U.-G. Mei{\ss}ner, Z.\ Phys.\ C \textbf{70}, 483 (1996).
\bibitem{Bernard:2007zu}V. Bernard, Prog.\ Part.\ Nucl.\ Phys.\ \textbf{60}, 82 (2008).
\bibitem{Borasoy:2005zg}B. Borasoy, P.C. Bruns, U.-G. Mei{\ss}ner, and R.~Nissler, Phys.\ Rev.\  C \textbf{72}, 065201 (2005).
\bibitem{Borasoy:2007ku}B. Borasoy, P.C. Bruns, U.-G. Mei{\ss}ner, and R. Nissler, Eur.\ Phys.\ J.\  A \textbf{34}, 161 (2007).
\bibitem{Ruic:2011wf}D. Ruic, M. Mai, and U.-G. Mei{\ss}ner, Phys. Lett. B \textbf{704}, 659 (2011).
\bibitem{DN10}M. D\"oring and K. Nakayama, Phys.\ Lett.\ B \textbf{683}, 145 (2010).
\bibitem{Kaiser:1996js}N. Kaiser, T. Waas, and W. Weise, Nucl.\ Phys.\  A \textbf{612}, 297 (1997).
\bibitem{Nacher:1999ni}J.C. Nacher, E. Oset, H. Toki, and A. Ramos, Phys.\ Lett.\ B \textbf{461}, 299 (1999).
\bibitem{Marco:1999df}E. Marco, S. Hirenzaki, E. Oset, and H. Toki, Phys.\ Lett.\ B \textbf{470}, 20 (1999).
\bibitem{MO00}J. Caro Ramon, N. Kaiser, S. Wetzel, and W. Weise, Nucl.\ Phys.\ A \textbf{672}, 249 (2000).
\bibitem{Meissner00}U.-G. Mei{\ss}ner and J.A. Oller, Nucl.\ Phys.\ A \textbf{673}, 311 (2000).
\bibitem{Gasparyan10}A. Gasparyan and M.F.M. Lutz, Nucl.\ Phys.\ A \textbf{848}, 126 (2010).
\bibitem{KSW95}N. Kaiser, P.B. Siegel, and W. Weise, Phys.\ Lett.\ B \textbf{362}, 23 (1995).
\bibitem{KL04}E.E. Kolomeitsev and M.F.M. Lutz, Phys.\ Lett.\ B \textbf{585}, 243 (2004).
\bibitem{SOV05}S. Sarkar, E. Oset, and M.J. Vincente Vacas, Nucl.\ Phys.\ A \textbf{750}, 294 (2005).
\bibitem{Doring:2007rz}M. D\"oring, Nucl.\ Phys.\ A \textbf{786}, 164 (2007).
\bibitem{Jido:2007sm}D. Jido, M. D\"oring, and E. Oset, Phys.\ Rev.\ C \textbf{77}, 065207 (2008).
\bibitem{Bruns:2010sv}P.C. Bruns, M. Mai, and U.-G. Mei{\ss}ner, Phys.\ Lett.\ B \textbf{697}, 254 (2011).
\bibitem{Ceci:2011ae}S. Ceci, M. D\"oring, C. Hanhart, S. Krewald, U.-G. Mei{\ss}ner, and A. \v Svarc, Phys.\ Rev.\ C \textbf{84}, 015205 (2011).
\bibitem{Schutz:1994ue}C. Sch\"utz, J.W. Durso, K. Holinde, and J. Speth, Phys.\ Rev.\ C \textbf{49}, 2671 (1994).
\bibitem{Schutz98}C. Sch\"utz, J. Haidenbauer, J. Speth, and J.W. Durso, Phys. Rev.\ C \textbf{57}, 1464 (1998).
\bibitem{Krehl00}O. Krehl, C. Hanhart, S. Krewald, and J. Speth, Phys.\ Rev.\ C \textbf{62}, 025207 (2000).
\bibitem{Gasparyan03}A.M. Gasparyan, J. Haidenbauer, C. Hanhart, and J. Speth, Phys.\ Rev.\ C \textbf{68}, 045207 (2003).
\bibitem{RDSHM10}M. D\"oring, C. Hanhart, F. Huang, S. Krewald, U.-G.
    Mei{\ss}ner, and D. R\"onchen, Nucl.\ Phys.\ A \textbf{851}, 58 (2011).
\bibitem{Doring:2011ip}M. D\"oring, J. Haidenbauer, U.-G. Mei{\ss}ner, and A. Rusetsky, arXiv: 1108.0676 [hep-lat].
\bibitem{Surya:1995ur}Y. Surya and F. Gross, Phys.\ Rev.\ C \textbf{53}, 2422 (1996).
\bibitem{Mainz}G.Y. Chen, S.S. Kamalov, S.N. Yang, D. Drechsel, and L. Tiator, Phys.\ Rev.\ C \textbf{76}, 035206 (2007).
\bibitem{Matsuyama07}A. Matsuyama, T. Sato, and T.-S.H. Lee, Phys.\ Rept.\
    \textbf{439}, 193 (2007) and references therein.
\bibitem{Diaz08}B. Juli\'a-D\'iaz, T.-S.H. Lee, A. Matsuyama, T. Sato, and L.C.
    Smith, Phys.\ Rev.\ C \textbf{77}, 045205 (2008).
\bibitem{Diaz09}B. Juli\'a-D\'iaz, H. Kamano, T.-S.H. Lee, A. Matsuyama, T. Sato, and N. Suzuki, Phys.\ Rev.\ C \textbf{80}, 025207 (2009).
\bibitem{Kamano09}H. Kamano, B. Juli\'a-D\'iaz, T.-S.H. Lee, A. Matsuyama, and T. Sato, Phys.\ Rev.\ C  \textbf{80}, 065203 (2009).
\bibitem{Feuster99} V. Shklyar, G. Penner, and U. Mosel, Eur.\ Phys.\ J. A \textbf{21}, 445 (2004).
\bibitem{Penner02}G. Penner and U. Mosel, Phys.\ Rev.\ C \textbf{66}, 055212 (2002).
\bibitem{KVI}A. Usov and O. Scholten, Phys.\ Rev.\ C \textbf{72}, 025205 (2005).
\bibitem{BonnGatchina}A.V. Sarantsev, V.A. Nikonov, A.V. Anisovich, E. Klempt, and U. Thoma, Eur.\ Phys.\ J. A \textbf{25}, 441 (2005).
\bibitem{Anisovich:2010an}A.V. Anisovich, E. Klempt, V.A. Nikonov, A.V. Sarantsev, and U. Thoma, Eur.\ Phys.\ J. A \textbf{47}, 27 (2010).
\bibitem{Drechsel:2007if}D. Drechsel, S.S. Kamalov, and L. Tiator, Eur.\ Phys.\ J. A \textbf{34}, 69 (2007).
\bibitem{Tiator:2011pw}L. Tiator, D. Drechsel, S.S. Kamalov, and M. Vanderhaeghen, arXiv: 1109.6745 [nucl-th].
\bibitem{Haberzettl06}H. Haberzettl, K. Nakayama, and S. Krewald, Phys.\ Rev.\ C \textbf{74}, 045202 (2006).
\bibitem{HHN2011}H. Haberzettl, F. Huang, and K. Nakayama, Phys.\ Rev.\ C \textbf{83}, 065502 (2011).
\bibitem{Schweber}For TOPT see, e.g., S.S. Schweber, ``An Introduction to Relativistic Quantum Field Theory'' (Harper\&Row, 1961; reprinted by Dover, 2005).
\bibitem{Wess67}J. Wess and B. Zumino, Phys.\ Rev.\ \textbf{163}, 1727 (1967).
\bibitem{Meissner:1987ge}U.-G. Mei{\ss}ner, Phys.\ Rept.\ \textbf{161}, 213 (1988).
\bibitem{Doring09L}M. D\"oring, C. Hanhart, F. Huang, S. Krewald, and U.-G. Mei{\ss}ner, Phys.\ Lett.\ B \textbf{681}, 26 (2009).
\bibitem{Doring09}M. D\"oring, C. Hanhart, F. Huang, S. Krewald, and U.-G. Mei{\ss}ner, Nucl.\ Phys.\ A \textbf{829}, 170 (2009).
\bibitem{Haberzettl97}H. Haberzettl, Phys.\ Rev.\ C \textbf{56}, 2041 (1997).
\bibitem{Kazes59}E. Kazes, Nuovo Cimento \textbf{13}, 1226 (1959).
\bibitem{Pascalutsa04}V. Pascalutsa and J.A. Tjon, Phys.\ Rev.\ C \textbf{70}, 035209 (2004).
\bibitem{Caia04}G.L. Caia, V. Pascalutsa, J.A. Tjon, and L.E. Wright, Phys. Rev.\ C \textbf{70}, 032201(R) (2004).
\bibitem{Caia05}G.L. Caia, L.E. Wright, and V. Pascalutsa, Phys.\ Rev.\ C \textbf{72}, 035203 (2005).
\bibitem{Gross}F. Gross and D.O. Riska, Phys.\ Rev.\ C \textbf{36}, 1928 (1987).
\bibitem{AA95}C.H.M. van Antwerpen and I.R. Afnan, Phys.\ Rev.\ C \textbf{52}, 554 (1995).
\bibitem{Kvinikhidze09}A.N. Kvinikhidze, B. Blankleider, E. Epelbaum, C. Hanhart, and M.P. Valderrama, Phys.\ Rev.\ C \textbf{80}, 044004 (2009).
\bibitem{WTI}J.C. Ward, Phys.\ Rev.\ \textbf{78}, 182 (1950); Y. Takahashi, Nuovo Cimento \textbf{6}, 370 (1957).
\bibitem{NH09}K. Nakayama and H. Haberzettl, Phys.\ Rev.\ C \textbf{80}, 051001(R) (2009).
\bibitem{HN10}H. Haberzettl and K. Nakayama, arXiv: 1011.1927 [nucl-th].
\bibitem{KVI02}H. Huisman \textit{et al.}, Phys.\ Rev.\ C \textbf{65}, 031001(R) (2002).
\bibitem{GellMann54}M. Gell-Mann and M.L. Goldberger, Phys.\ Rev.\ \textbf{96}, 1433 (1954).
\bibitem{BallChiu1980}J.S. Ball and T.W. Chiu, Phys.\ Rev.\ D \textbf{22}, 2542 (1980).
\bibitem{BbS}V.G. Kadyshevsky, Nucl.\ Phys.\ B \textbf{6}, 125 (1968).
\bibitem{Janssen96}G. Janssen, K. Holinde, and J. Speth, Phys.\ Rev.\ C \textbf{54}, 2218 (1996).
\bibitem{Dugger09}M. Dugger {\it et al.}, CLAS Collaboration, Phys.\ Rev.\ C \textbf{79}, 065206 (2009).
\bibitem{Ahrens04}J. Ahrens {\it et al.}, GDH and A2 Collaboration, Eur.\ Phys.\ J. A \textbf{21}, 323 (2004).
\bibitem{SAID}CNS Data Analysis Center, The George Washington University, \url{http://gwdac.phys.gwu.edu/}.
\bibitem{Bartalini05}O. Bartalini {\it et al.}, GRAAL Collaboration, Eur.\ Phys.\ J. A \textbf{26}, 399 (2005).
\bibitem{Bartholomy05}O. Bartholomy {\it et al.}, CB-ELSA Collaboration, Phys.\ Rev.\ Lett.\ \textbf{94}, 012003 (2005).
\bibitem{Shafi04}A. Shafi {\it et al.}, Crystal Ball Collaboration, Phys.\ Rev.\ C \textbf{70}, 035204 (2004).
\bibitem{Bacci72}C. Bacci {\it et al.}, Phys. Lett. B \textbf{39}, 559 (1972).
\bibitem{Hemmi73}Y. Hemmi {\it et al.}, Nucl. Phys. B \textbf{55}, 333 (1973).
\bibitem{Elsner09}D. Elsner {\it et al.}, CB-ELSA Collaboration and TAPS Collaboration, Eur.\ Phys.\ J. A \textbf{39}, 373 (2009).
\bibitem{Salvo09}R. Di Salvo {\it et al.}, GRAAL Collaboration, Eur. Phys. J. A \textbf{42}, 151 (2009).
\bibitem{BKLM94}V. Bernard, N. Kaiser, T.-S.H. Lee, and U.-G. Mei\ss ner, Phys.\ Rept.\  \textbf{246}, 315 (1994).
\bibitem{Korkmaz99}E.J. Korkmaz {\it et al.}, Phys.\ Rev.\ Lett.\ \textbf{83}, 3609 (1999).
\bibitem{Schmidt01}A. Schmidt {\it et al.}, Phys.\ Rev.\ Lett.\ \textbf{87}, 232501 (2001).
\bibitem{Fuchs96}M. Fuchs {\it et al.}, Phys.\ Lett.\ B \textbf{368}, 20 (1996).
\bibitem{DN10a}M. D\"oring and K. Nakayama, Eur.\ Phys.\ J. A \textbf{43}, 83 (2010).
\bibitem{Tabakin}W.T. Chiang and F. Tabakin, Phys.\ Rev.\ C \textbf{55}, 2054 (1997).
\bibitem{Tiator11}L. Tiator, arXiv: 1109.0608 [nucl-th].
\bibitem{Workman10}R.L. Workman, Phys.\ Rev.\ C \textbf{83}, 035201 (2011).
\bibitem{Janssen94}G. Janssen, K. Holinde, and J. Speth, Phys.\ Rev.\ Lett.\ \textbf{73}, 1332 (1994).
\bibitem{Janssen942}G. Janssen, K. Holinde, and J. Speth, Phys.\ Rev.\ C \textbf{49}, 2763 (1994).
\bibitem{Machleidt87}R. Machleidt, K. Holinde, and Ch. Elster, Phys.\ Rept. \textbf{149}, 1 (1987).
\bibitem{NDHHS99}K. Nakayama, J.W. Durso, J. Haidenbauer, C. Hanhart, and J. Speth, Phys.\ Rev.\ C \textbf{60}, 055209 (1999).
\bibitem{PDG}Particle Data Group, J. Phys.\ G \textbf{37}, 075021 (2010).
\bibitem{Garcilazo93}H. Garcilazo and E. Moya de Guerra, Nucl.\ Phys.\ A \textbf{562}, 521 (1993).
\bibitem{NOH08}K. Nakayama, Yongseok Oh, and H. Haberzettl, J. Kor.\ Phys.\ Soc.\ \textbf{59}, 224 (2011) [arXiv: 0803.3169 [hep-ph]].
\end{thebibliography}
\end{document}